\documentclass[12pt,preprint]{aastex}

\usepackage{bm}
\usepackage{epstopdf} 
\usepackage{natbib}
\usepackage[backref,colorlinks,citecolor=blue]{hyperref}
\usepackage{threeparttable}
\usepackage{booktabs}
\usepackage{graphicx}
\usepackage{gensymb}

\shorttitle{hot Extreme Helium Stars} 
\shortauthors{Anirban, Pandey, Lambert}

\begin{document}
\title{Detection of fluorine in hot Extreme Helium Stars}

\author{Anirban Bhowmick$^{1}$, Gajendra Pandey$^{1}$, David L. Lambert$^{2}$}
\affil{$^{1}$Indian Institute of Astrophysics, Koramangala, Bengaluru
 560 034, India}
\affil{$^{2}$ W.J. McDonald Observatory and Department of Astronomy, The University of Texas at Austin, Austin, TX 78712, USA}

\email{anirban@iiap.res.in; pandey@iiap.res.in; dll@astro.as.utexas.edu}

\begin{abstract}

The main objective of this paper is to explore abundances of fluorine in hot Extreme Helium Stars (EHes). Overabundance of fluorine is a characteristic feature for cool EHes and R Coronae Borealis (RCB) stars and further enforces their close connection. For hot EHes this relationship with the cooler EHes, based on their fluorine abundance is unexplored. We present in this paper the first  abundance estimates of fluorine determined from singly ionised fluorine lines (F\,{\sc ii}) for 10 hot EHe stars from optical spectra. Fluorine abundances were determined using the F\,{\sc ii} lines in two windows centered at 3505  \AA \ and 3850 \AA  . Six of the 10 stars show significant enhancement of fluorine similar to the cool EHes. Two carbon-poor hot EHes show no signature of fluorine and have a significant low upper limit for the F abundance. These fluorine abundances are compared with the other elemental abundances observed in these stars which provide an idea about the formation and evolution of these stars. The trends of fluorine with C, O, and Ne show that significant helium burning after a CO-He white dwarf merger can account for a majority of the observed abundances. Predictions from simulations of white dwarf mergers are discussed in light of the observed abundances.
\end{abstract}

\section{INTRODUCTION}

Extreme helium stars (EHes) are helium rich, hydrogen deficient A and B-type supergiants having effective temperatures in the range of 8000$-$35000 K. The observed surface composition of these stars are  similar to the cooler hydrogen deficient stars, namely the  R Coronae Borealis (RCB) and hydrogen deficient carbon (HdC) stars. Apart from sharing extreme hydrogen deficiency, EHe, RCB and HdC stars also exhibit common peculiar aspects of their chemical compositions. 

The two most notable peculiarities in the compositions of these H-deficient stars are (i) the extreme overabundance of $^{18}$O in HdC and cool RCBs such that $^{18}$O/$^{16}$O $> 1$ \citep{Clayton2007} and (ii) a startling overabundance of F in RCBs and cool EHes such that F relative to Fe is enhanced by 800 to 8000 times \citep{Pandey2006f,Pandey2008,Hema2017}. It is now of great interest to determine, as seems plausible, if these peculiarities extend to the hot EHes. This paper addresses the F abundance of the hot EHes. 

EHes are rare in the Galaxy and hot EHes are necessarily extremely rare. \citet{Jeffery1996} list 21 EHes in their catalogue. An additional EHe was reported recently \citep{Jeffery2017}. There are about 17 known hot EHes with effective temperatures hotter than about  14000 K, the focus of this paper. Ten hot EHes are examined here. For the hot EHes, nothing is known about the two notable abundance anomalies of the H-deficient cool stars, i.e.,  $^{18}$O and F. Since the O isotopic abundances are determined from CO lines in the infrared spectrum and CO molecules cannot exist in the atmospheres of hot (or cool) EHes, the O isotopic abundances are unobtainable for EHes. (Isotopic wavelength shifts for O\,{\sc i} and O\,{\sc ii} lines are negligible.) Fluorine abundances are, however, obtainable for EHes.

The chemical compositions derived from their observed spectra suggest a hydrogen-deficient atmosphere including  material exposed to both H- and He-burning. Based on their observed surface compositions two principal theories are in place to explain their origins: the ``double-degenerate" (DD) model and the ``final-flash" (FF) model. Based primarily on the fluorine, neon, $^{13}$C, and $^{18}$O abundances, a consensus is now emerging for the DD scenario, however, a small fraction may be produced by FF scenario. The principal version of the DD model involves the merger of an He white dwarf with a more massive C-O white dwarf following the decay of their orbit \citep{Webbink1984,Iben1984}. Other mergers may involve two He white dwarfs. The second model, the FF model, refers to a late or final He-shell flash in a post-AGB star. In this model \citep{Iben1983}, the ignition of the helium shell in a post-AGB star, say, a cooling white dwarf, results in what is known as a late or very late thermal pulse \citep{Herwig2001}. 

Simulations predict that a CO-He white dwarf merger in the DD scenario may produce  conditions for partial helium burning  which results in production of $^{18}$O via  $^{14}$N$(\alpha, \gamma)^{18}$F$(\beta^{+}\nu)^{18}$O and of $^{19}$F, the sole stable isotope of F \citep{Clayton2007}.   Hence, the knowledge of the fluorine abundance  and its relation to the other abundant species found in these stars plays an important role in discovering the nucleosynthesis processes taking place during and following helium accretion on to the C-O or He white dwarf in the DD scenario.

If the suite of abundance peculiarities is seen to be common across the HdC, RCB and EHe, primarily a sequence of increasing effective temperature, a common formation scenario would seem a likely scenario. As noted above, the $^{18}$O anomaly cannot be investigated in EHes. The F anomaly is determinable across the sequence. For warm RCBs and the cooler EHes, neutral fluorine (F\,{\sc i}) lines have provided the high F overabundances \citep{Pandey2006f, Pandey2008, Hema2017}. For hot EHes, the F\,{\sc i} lines are undetectable in optical spectra but lines of ionized fluorine should be present in ultraviolet (3500-3900 \AA) spectra if the F abundance is anomalous. To date, the only confirmed detection of F\,{\sc ii} lines in a H-deficient star is \citet{Pandey2014}'s detection of F\,{\sc ii} lines at 3500-3510 \AA \ in a spectrum of the hot EHe/hot RCB DY Cen. However, DY Cen is an odd H-deficient star in that it has a relatively high hydrogen abundance. Detection of fluorine in other hot EHes has yet to be explored. Here we report F abundances (or upper limits) for ten hot EHes.

The paper is organized as follows: Section 2 discusses the observations, Section 3 addresses the identification of the F\,{\sc ii} lines, Section 4 presents the abundance analysis and discusses the relations between the F and some other elemental abundances. Section 5 discusses  the compositions of the hot EHes and other H-deficient stars in the light of predictions from simulations of the DD scenario. Section 6 concludes the paper with a few final remarks.

\section{Observation}

High resolution optical echelle spectra of ten hot EHes come from HCT-HESP, ESO-FEROS and ESO-UVES, and McDonald Observatory, as discussed below. All but two stars (DY\,Cen and V1920\,Cyg) were observed with more than one telescope/spectrograph combination (see Table \ref{Table.1}).

We  observed three hot EHes: V652\,Her, V2205\,Oph and BD +$10^{\degree}$\,2179 using Hanle Echelle Spectrograph (HESP) \citep{Sriram2018} mounted on the 2-m Himalayana Chandra Telescope (HCT) at the Indian Astronomical Observatory (IAO) in Hanle, Ladakh, India  during 2017 and 2018 to look specifically  for F\,{\sc ii} lines in the 3500\AA \ and 3800\AA \ regions. The observing details are in Table \ref{Table.1}. A Th-Ar lamp was observed  for wavelength calibration. To normalise the pixel-to-pixel variation in the sensitivity of the CCD, many exposures known as flat frames with differing spectrograph focus (in focus and out of focus) were obtained using a featureless quartz-halogen lamp. All the flat frames were combined to create a master flat with very high signal  for flat correction. A spectrum of a rapidly rotating B-type bright star was obtained during each observing run  for aperture extraction of faint programme stars, and also for removing the atmospheric lines. The data was reduced using standard IRAF\footnote{IRAF is distributed by the National Optical Astronomy Observatory, which is operated by the Association of Universities for Research in Astronomy (AURA) under a cooperative agreement with the National Science Foundation.} packages for bias correction, flat correction, aperture extraction and wavelength calibration. The final wavelength calibrated spectra of these three stars V652\,Her, V2205\,Oph and BD +$10^{\degree}$\,2179 were combined (see below) with spectra from the ESO Data Archives. 

We also retrieved high resolution optical spectra of ten hot EHes  from the European Southern Observatory (ESO) Data Archives\footnote{\url{http://archive.eso.org/wdb/wdb/adp/phase3_main/form}}. These observations were made with ESO Telescopes at the La Silla and the Paranal Observatory under programme IDs 077.D$-$0458, 284.D$-$5048, and 074B$-$0455. The spectra were recorded using FEROS on ESO 2.2m telescope in La Silla, Chile and UVES on ESO Very Large Telescope (VLT) at Paranal, Chile. The details are given  in Table \ref{Table.1}. FEROS  provides the useful wavelength range of 3530 \AA \ to 9200 \AA, whereas the UVES  provides spectra in following wavelength windows: 3050-3870 \AA, 3280-4560 \AA, 5655-9460 \AA,  6650-8540 \AA \ and 8650-10240 \AA .

The spectrum of hot EHe star V1920\,Cyg was observed using the W.J. McDonald Observatory's Harlan J. Smith 2.7-m telescope with the Robert G. Tull cross-dispersed echelle spectrograph during 1996 at a resolving power of about 30,000 \citep{Tull1995}. V1920\,Cyg's spectrum is discussed in \citep{Pandey2006}, and the relevant details are also provided in  Table \ref{Table.1}).

Spectra retrieved from archival data and those obtained from HESP were further smoothed to increase the signal-to-noise ratio.  These spectra were finally normalised to continuum. Note that the spectra were smoothed to the limit that the stellar line profiles remain unaltered. To ensure this, the smoothed spectrum was compared with the unsmoothed one. The resolving power of the smoothed spectrum was determined by measuring the FWHM of telluric lines in the 6925\AA \ region. If telluric lines were not available for determining the spectral resolution of the smoothed spectra, the reported resolving power in the archives was used by taking into account the smoothening factor.

Frames with symmetric absorption line profiles and with minimum core emission were chosen for analysis; many EHes show variable spectra with radial velocity changes, variable line profiles and even emission features. The spectra obtained from each individual frames were compared to check for the presence of any artifact. The signal in the spectra obtained through HESP was very low in 3500\AA \ region, hence, we have used only the spectral region above 3800\AA \ region for analysis. To further improve the signal-to-noise, the spectrum from archival data and that from HESP, if available, were co-added for final analysis. Note that the observed spectra are brought to the rest wavelength using well known stellar lines. The details of the final co-added spectra are given in Table \ref{Table.2}.

\begin{table*}[ht]

\caption{Log of observations of the EHe stars.}
\begin{threeparttable}
\resizebox{\textwidth}{!}{%
\begin{tabular}{llllllll}\hline\label{Table.1}
Star name  & Date of observation & Exposure & $V$-mag  & S/N (3500 \AA) & S/N(3800 \AA) & Source of & $R=\lambda/\Delta\lambda$ \\ 
 &  & time(secs) & & & & spectra & \\

\hline
LS\,IV+$6^{\degree}$ 2 & 2006-03-31 & 2000 & 12.2 & 120 & & UVES & 40000 \\
 & 2006-04-21 & 2980 & 12.2 & & 175 & FEROS & 45000\\
 
V652\,Her & 2005-03-01 & 600 & 10.5 & & 110 & FEROS & 45000\\
 & 2017-06-04  & 2700(5) & 10.5 & & 65 & HESP & 28000\\
 & 2018-04-22  & 2700(3) & 10.5 & & 40 & HESP & 28000\\

DY\,Cen & 2010-02-27 & 1800 & 12.5 & 140 & 120 & UVES & 40000\\ 

V2205\,Oph & 2005-02-26 & 600 & 10.5 & & 100 & FEROS & 45000\\
 & 2017-06-04 & 2400(4) & 10.5 &  & 60 & HESP & 28000\\
 & 2018-05-09 & 2400(3) & 10.5 &  & 50 & HESP & 29000\\
 & 2018-05-10 & 2400(3) & 10.5 &  & 38 & HESP & 29000\\
 
HD\,144941 & 2006-04-10 & 780 & 10.1 & 270 & & UVES & 40000\\
 & 2006-01-08 & 3000 & 10.1 & & 250 & FEROS & 45000\\
 
LSE\,78 & 2006-01-10 & 1500 & 11.2 & 155 & & UVES & 40000\\
 & 2006-04-09 & 2400 & 11.2 & & 170 & FEROS & 45000\\
 
BD +$10^{\degree}$\,2179 & 2006-05-10 & 1000 & 10.0 & 220 & & UVES & 40000 \\
 & 2006-04-12 & 2820 & 10.0 & & 210 & FEROS & 45000\\
 & 2018-01-13 & 2400(3) & 10.0 & & 95 & HESP & 29000\\ 
 & 2018-02-10 & 2400(3) & 10.0  & & 110 & HESP & 29000\\ 
 & 2018-03-27 & 2400(3) & 10.0 & & 80 & HESP & 29000\\

V1920\,Cyg & 1996-07-25 & 1800 & 10.3 & & 110 & McDonald & 48000\\

HD\,124448 & 2006-04-10 & 975 & 10.0 & 190 & & UVES & 40000\\
   & 2006-04-08 & 2820 & 10.0 & & 200 & FEROS & 45000\\ 
   
PV\,Tel  & 2006-04-08 &  1500 & 9.3 & & 180 & FEROS & 45000\\ \hline \hline
\end{tabular}}

\end{threeparttable}
\end{table*}

\begin{table*}[ht]
\caption{Details of the spectra}
\begin{center}
\begin{tabular}{lllllll}\hline\label{Table.2}
Star name  & \multicolumn{5}{c}{Wavelength window}\\  
\cmidrule{2-6} 
 &  & \multicolumn{2}{c}{3505\AA}  & & \multicolumn{2}{c}{3850\AA} \\ 
\cmidrule{2-3} \cmidrule{5-6}
 &  S/N & R($\lambda$/$\Delta\lambda$) & & S/N & R($\lambda$/$\Delta\lambda$) \\
\midrule \hline
LS\,IV+$6^{\degree}$ 2  & 225 & 35000 & & 260 & 37500 \\ 
V652\,Her  & \nodata & \nodata & & 175 & 26000 \\ 
DY\,Cen  & 160 & 33000 & & 210 & 31000 \\
V2205\,Oph  & \nodata & \nodata & & 320 & 27000 \\ 
HD\,144941   & 370 & 38000 & & 340 & 37000 \\
LSE\,78   & 280 & 36000 & & 220 & 36000 \\
BD +$10^{\degree}$\,2179  & 320 &  38000 &  & 280 & 28000 \\ 
V1920\,Cyg   & \nodata & \nodata & & 140 & 30000\\
HD\,124448    & 220 & 39000 & & 240 & 37500 \\ 
PV\,Tel & \nodata & \nodata & & 220 & 38000 \\ \hline 

 \end{tabular}
\end{center}
\end{table*}

\begin{table*}[!ht]
\caption{F\,{\sc ii} lines from 3s $-$ 3p and 3p $-$ 3d transition array contributing to the spectra of the analysed stars. The F\,{\sc ii} lines used in abundance determinations are shown in bold.}
\resizebox{\textwidth}{!}{%
\begin{tabular}{c c c c l} \hline \label{Table.3}
 Multiplet No. & $\lambda$ & $\chi$ & log \textit{gf} & Likely contributors \\
  & \AA & (ev) & &  \\ \hline
1 & \textbf{3847.086} & 21.88 & 0.31 & F\,{\sc ii}, N\,{\sc ii} $\lambda$ 3847.38  \\ 
  & \textbf{3849.986} & 21.88 & 0.16 & F\,{\sc ii}, Mg\,{\sc ii}(weak) $\lambda$ 3850.40 \\
  & \textbf{3851.667} & 21.88 & $-$0.06 & F\,{\sc ii}, O\,{\sc ii} $\lambda$ 3851.47  \\ 
  & & & & \\
2 & 4024.727 & 22.67 & 0.16 & F\,{\sc ii} , He\,{\sc i} , $\lambda$ 4023.986 , 4026.189,  4026.362  (very strong) \\
  & 4025.010 & 22.67 & $-$0.54 & F\,{\sc ii} , He\,{\sc i} , $\lambda$ 4023.986 , 4026.189,  4026.362  (very strong) \\
  & 4025.495 & 22.67 & $-$0.06 & F\,{\sc ii} , He\,{\sc i} , $\lambda$ 4023.986 , 4026.189,  4026.362  (very strong) \\
  & & & & \\
3 & \textbf{3505.614} & 25.10 & 0.676 & F\,{\sc ii} \\
  & \textbf{3505.520} & 25.10 & 0.09 & F\,{\sc ii} \\
  & \textbf{3505.370} & 25.10 & $-$0.757 & F\,{\sc ii} \\
  & 3503.095 & 25.10 & 0.391 & F\,{\sc ii}, Ne\,{\sc ii} $\lambda$ 3503.61 \\
  & 3502.954 & 25.10 & 0.187 & F\,{\sc ii}, He\,{\sc i} $\lambda$ 3502.393 (strong) \\
  & 3501.416 & 25.10 & 0.074 & F\,{\sc ii}, He\,{\sc i} $\lambda$ 3498.659 (very strong) ,  Fe\,{\sc iii} $\lambda$ 3501.767 \\
  & & & & \\
4 & 4103.525 & 25.75 & 0.559 & F\,{\sc ii}, O\,{\sc ii} $\lambda$ 4103.017, N\,{\sc iii} $\lambda$ 4103.37 (strong) \\
  & 4103.085 & 25.75 & 0.289 & F\,{\sc ii}, O\,{\sc ii} $\lambda$ 4103.017, N\,{\sc iii} $\lambda$ 4103.37 (strong) \\
  & 4103.724 & 25.75 & $-$0.064 &  F\,{\sc ii},  N\,{\sc iii} $\lambda$ 4103.37 (strong) \\
  & 4103.871 & 25.75 & $-$0.19 & F\,{\sc ii} N\,{\sc iii} $\lambda$ 4103.37 (strong) \\
& & & & \\
5 & 4109.173 & 26.26 & 0.45 & F\,{\sc ii}, O\,{\sc ii} $\lambda$ 4108.75  , Mg II, $\lambda$ 4109.54 \\
  & 4116.547 & 26.27 & 0.18 & F\,{\sc ii}, Si\,{\sc iv} $\lambda$ 4116.104 (strong) \\
  & 4119.219 & 26.27 & $-$0.01 & F\,{\sc ii}, O\,{\sc ii} $\lambda$ 4119.221 (strong) \\ \hline
    
\end{tabular}}

\end{table*}

\section{Identification of F\,{\sc ii} lines}

Multiplets numbered 1, 2, 3, 4, and 5 in the  Revised Multiplet Table of \citet{Moore1972} and by \citet{Wiese1966} are the potential contributors of F\,{\sc ii} absorption lines to the spectra of hot EHe stars. A complete list of the transitions that includes their wavelengths, lower excitation potential, and log-$gf$ values for lines of these multiplets was compiled from the NIST database \footnote{\url{https://physics.nist.gov/PhysRefData/ASD/lines_form.html}}.

\begin{figure*}
\center
\includegraphics[scale=0.5]{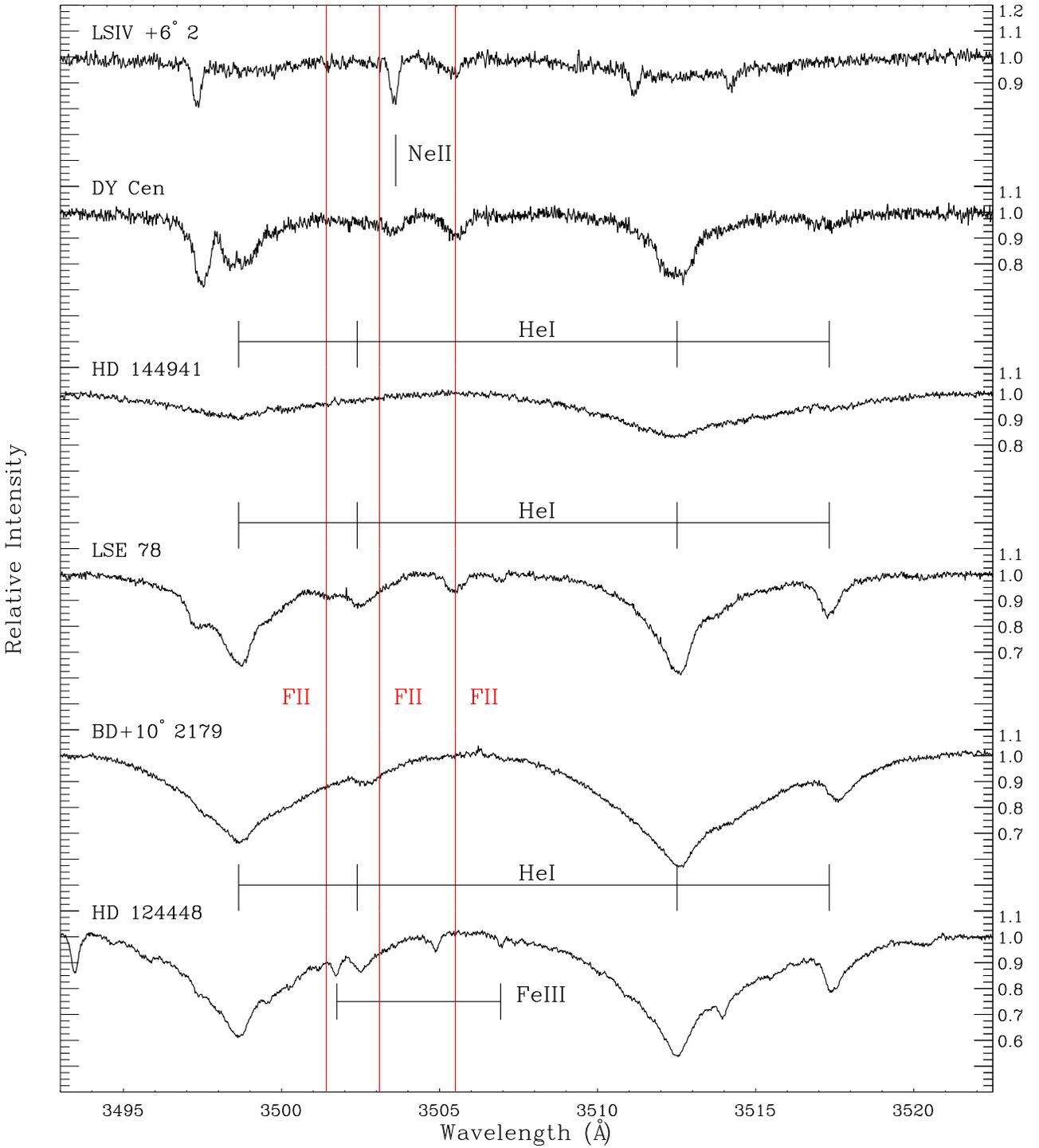}
\caption{\small{Comparison of the spectra with key identifications in 3500 \AA $ $ region. The stars are arranged according to their effective temperature with hottest on the top and coolest at the bottom. The red lines represents the F\,{\sc ii} lines of RMT 3 in this window. }}\label{fig.1}
\end{figure*}

\begin{figure*}
\center
\includegraphics[scale=0.4]{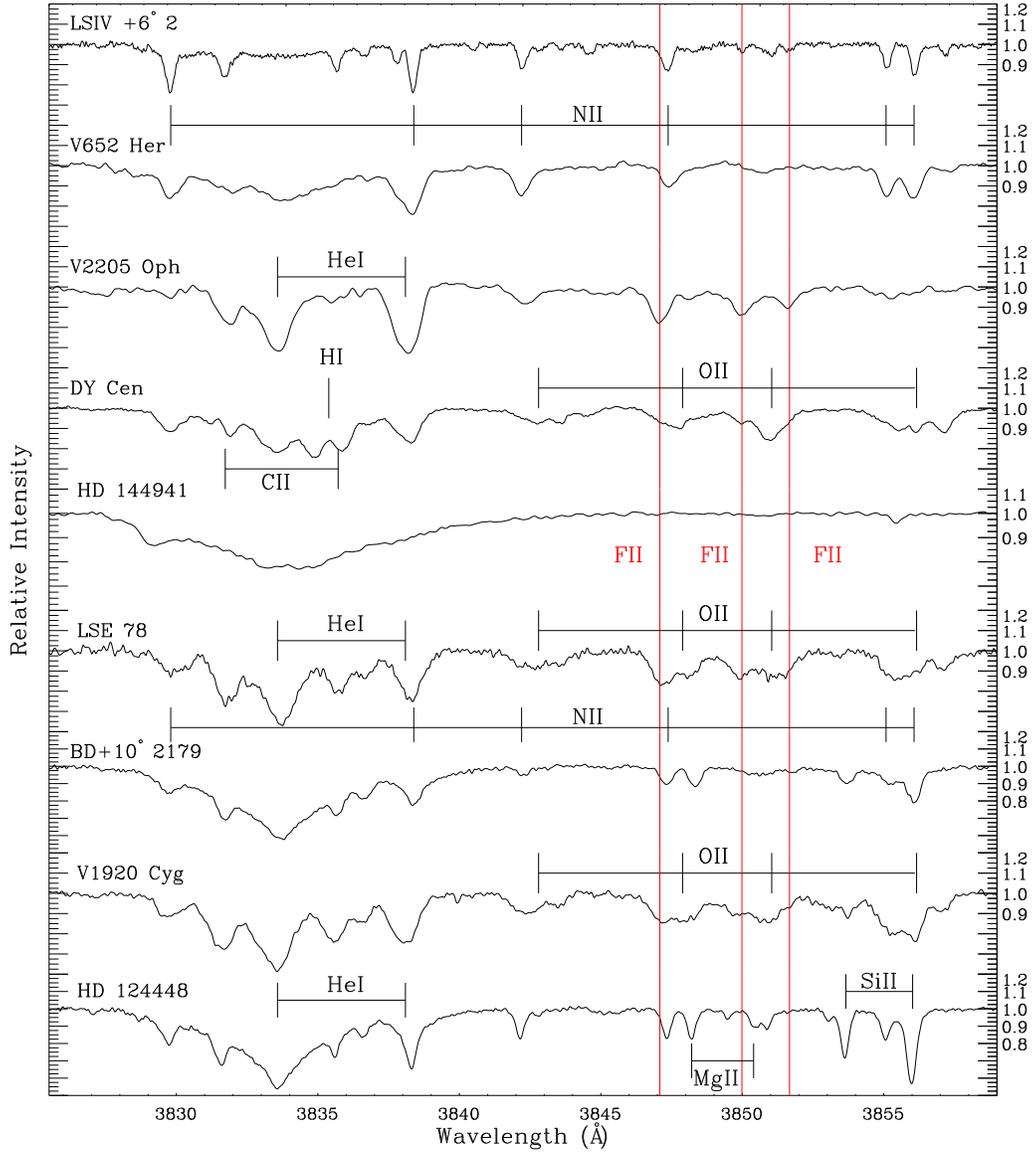}
\caption{\small{Comparison of the spectra with key identifications in 3850 \AA $ $ region. The stars are arranged according to their effective temperature with hottest on the top and coolest at the bottom. The red lines represents the F\,{\sc ii} lines of RMT 1 in this window.}}\label{fig.2}
\end{figure*}

Four F\,{\sc ii} lines were identified as the main or significant contributor to  stellar lines (see Table \ref{Table.3}. These four lines consist of all three lines of muliplet 1$-$ 3847.086\AA , 3849.986 \AA \ and 3851.667 \AA \ and the fourth line centered at 3505.614 \AA \ of multiplet 3. Note that the F\,{\sc ii} line profile at 3505.614 \AA \ which appears as one,  is a blend of 3 components 3505.614 \AA \, 3505.52 \AA \ and 3505.37 \AA \ (see Table \ref{Table.3}).   Lines at 3849.986 \AA \ of multiplet 1 and 3505.614  \AA \  of multiplet 3 are relatively free of blends and are best suited for determining the F abundance (see Table \ref{Table.3}). All the lines of multiplets 1 and 3 are shown in Figures \ref{fig.1} and \ref{fig.2}, where the spectra of hot EHes are ordered from top to bottom in order of decreasing effective temperature. The wavelength windows corresponding to Figures \ref{fig.1} and \ref{fig.2}  are centered around 3508 \AA \ and 3850 \AA \ , respectively. Note that the spectra of V652\,Her, V2205\,Oph, and V1920\,Cyg were not available or were very noisy in the window 3490-3520\AA . Also for the other multiplets of F\,{\sc ii} lines, a thorough search was conducted for the blending lines and strong blending of lines from other atomic species is noted (see Table \ref{Table.3}). These multiplets were not selected for measuring the fluorine abundance: multiplet 2 is heavily blended with a Stark broadened strong He\,{\sc i} line profile, multiplet 4 and 5 are blended severely by lines of other elements.
The blended lines were identified using the Revised Multiplet Table \citep{Moore1972}, Tables of spectra of H, C, N, and O \citep{Moore1993}, and the NIST Atomic Spectra Database \footnote{\url{https://physics.nist.gov/PhysRefData/ASD/lines_form.html}} that also provides the line's $gf$-value. 

\section{Abundance analysis}

The abundance of an element X in normal stars is quoted with respect to hydrogen (i.e., X/H) due to hydrogen being the main contributor to the continuous opacity directly or indirectly as well as the most abundant element in their atmospheres. A measure of fractional abundance for the element X is also given by the mass fraction, Z(X) where

\begin{equation}
Z(X)=\frac{\mu_X N_X}{\mu_H N_H + \mu_{He} N_{He} + ...+\mu_i N_i}
= \frac{\mu_X N_X}{(\sum\mu_i N_i)}
\end{equation}

\begin{equation}
\simeq\frac{\mu_X A_X}{1 + 4 A_{He}}
\end{equation}

where $\mu_{X}$ is the atomic weight of element X, and A$_{X}$ = X/H. Hence, Z(X) is directly estimated from A$_{X}$, a result of abundance analysis of the observed spectrum, and an assumption about A$_{He}$ if helium lines are not observable. 

For the case of hydrogen-poor stars like the hot EHes, helium may be the main contributor to the continuum opacity directly or indirectly and also the most abundant element in their atmospheres where the H/He ratio has been changed by the addition of nuclear-processed material from H- and He-burning layers. Thus, the abundance of an element X is determined with respect to helium i.e., X/He = A$'_{X}$, the equation 1 reduces to

\begin{equation}
Z(X)= \frac{\mu_X A'_X}{H/He + 4 + 12C/He + ..\mu_iA'_i }
\end{equation}

Due to hydrogen being very poor in these stars, H/He is very small and like other trace elements can be ignored, then the above equation reduces to,

\begin{equation}
Z(X) \simeq\frac{\mu_X A'_X}{4 + 12C/He}
\end{equation}

The C/He can be spectroscopically determined for hot EHes ($\simeq$ 0.01), and also the abundance of any element X for a hot hydrogen-deficient star like hot EHes, can be directly measured spectroscopically i.e., A$'_{X}$ = X/He.

Due to the conservation of nucleons during different stages of nuclear burning, the derived abundances are normalised based on the convention that $\log\epsilon$(X) = $\log(X/H)$ + 12.0 to a scale in which $\log\sum\mu_i\epsilon(i)$ = 12.15, where 12.15 is determined from solar abundances with  He/H $\simeq$ 0.1. Based on this normalisation convention, and considering X/He as the
measure of abundance of an element X in hot H-poor or hot EHe stars, the helium abundance $\log\epsilon$(He) is about 11.54 from equation 4.

The F abundance is derived from the four best F\,{\sc ii} lines (Table \ref{Table.3}). Since these lines are subject to blending, spectrum synthesis was used to locate a F\,{\sc ii} line's contribution. The code SYNSPEC \citep{Hubeny1994} was used with the LTE model atmospheres of individual stars (see Table \ref{Table.4}) from \citet{Pandey2006,Pandey2011,Pandey2014,Pandey2017}. Synthetic spectra were convolved with the instrumental profile and the broadening corresponding to the rotational velocity derived from  weak and symmetric O\,{\sc ii} or N\,{\sc ii} lines in the star's spectrum. All the key lines  were used to compose a line list for spectrum synthesis. Selected lines of several elements were synthesized. Derived LTE abundances are in fair agreement with those reported in our earlier abundance analyses \citep{Pandey2006,Pandey2011,Pandey2014,Pandey2017}. Adopted model atmospheres (T$_{eff}$ = effective temperature, log$g$ = surface gravity, $\xi$ = microturbulence) and the F abundances from the individual F\,{\sc ii} lines and the line-to-line scatter are given in Table \ref{Table.4}. Abundances of other elements (C, N, O, Ne, Fe and Zr) are given in Table \ref{Table.5}. 

The two spectral regions providing the F\,{\sc ii} lines are displayed in Figures \ref{fig.1} and \ref{fig.2} with the EHes arranged in order of decreasing effective temperature from top to bottom. By inspection, it is obvious that the F\,{\sc ii} lines do not vary smoothly with effective temperature; the F abundance can be greatly different in stars of similar effective temperature. Consider, for example, V652\,Her and V2205\, Oph in Figure \ref{fig.2} with the three F\,{\sc ii} lines prominent in the spectrum of V2205\,Oph but seemingly absent from the spectrum of V652\,Her. The two stars have similar atmospheric parameters  but F abundances differing by at least 1.5 dex (Table \ref{Table.4}).

\begin{table*}[ht]
\caption{Derived abundances of fluorine in  hot EHes.}
\begin{center}

\begin{threeparttable}
\resizebox{\textwidth}{!}{%
\begin{tabular}{llrrrrrll}\hline\label{Table.4}
Star name & ($T_{eff}$, log\textit{g}, $\xi$)& \multicolumn{6}{c}{log$\epsilon$(F)}\\  
\cmidrule{3-9} 
 &  & 3847.086 \AA & 3849.986 \AA & 3851.667 \AA & 3505.614 \AA & Mean & $\sigma_{1}$\tnote{a} & $\sigma_{2}$\tnote{b} \\ 

\midrule
LS\,IV+$6^{\degree}$ 2   & (32000, 4.20, 9.0)\tnote{1} & 6.5 & 6.4 & 6.4 & 6.6 & 6.5 & 0.1 & $\pm$0.1 \\ 
V652\,Her & (25300, 3.25, 13.0)\tnote{2} & $<$ 5.7 & $<$ 5.5 & $<$ 5.6 &\nodata & $<$ 5.6 & \nodata &\nodata\\ 
V2205\,Oph  & (24800, 2.85, 23.0)\tnote{1} & 7.0 & 7.0 & 7.0 &\nodata & 7.0 &  0.1 & $\pm$ 0.1 \\ 
DY\,Cen  & (24750, 2.65, 24.0)\tnote{4} & 6.7 & 6.9 & 6.8 & 7.0 & 6.9 &  0.1 & $\pm$ 0.2 \\ 
HD\,144941  & (21000, 3.35, 10.0)\tnote{2} & $<$ 5.5 & $<$ 5.7 & $<$ 5.5 & $<$ 5.5 & $<$ 5.6 & \nodata &\nodata\\ 
LSE\,78  & (18300, 2.2, 16.0)\tnote{4} & 7.4 & 7.4 & 7.4 & 7.3 & 7.4 &  0.1 & $\pm$ 0.2\\ 
BD +$10^{\degree}$\,2179  & (17000, 2.6, 7.5)\tnote{1} & 6.4 & 6.5 & 6.4 & $<$ 6.5 & 6.4 &  0.2 & $\pm$ 0.1\\ 
V1920\,Cyg  & (16300, 1.8, 20)\tnote{4} & 7.5 & 7.6 & 7.5 &\nodata & 7.5 &  0.2 & $\pm$ 0.1\\ 
HD\,124448  & (15500, 2.0, 12)\tnote{4} & $<$ 6.0 & $<$ 6.0 & $<$ 6.0 & $<$ 6.0 & $<$ 6.0 & \nodata &\nodata\\ 
PV\,Tel     & (13750, 1.6, 25.0)\tnote{1} & $<$ 6.5 & $<$ 6.5 & $<$ 6.5 & \nodata & $<$ 6.5 & \nodata &\nodata\\ \hline
\end{tabular}}
\begin{tablenotes}
\item [a] r.m.s error: $\Delta T_{eff}$ = $\pm$ 500K , $\Delta$log\textit{g} = $\pm$ 0.2 cgs
\item [b] r.m.s error: line-to-line scatter
\item [1] \citep{Pandey2011}
\item [2] \citep{Pandey2017}
\item [3] \citep{Pandey2014}
\item [4] \citep{Pandey2006}

\end{tablenotes}
\end{threeparttable}
\end{center}
\end{table*}

Brief remarks follow on the spectrum syntheses of the F\,{\sc ii} lines in the individual stars and their F abundances beginning with the hottest star LS\,IV+$6^{\degree}$ 002.  Observed and synthetic spectra are shown for each star.

\begin{figure*}
\center
\includegraphics[scale=0.5]{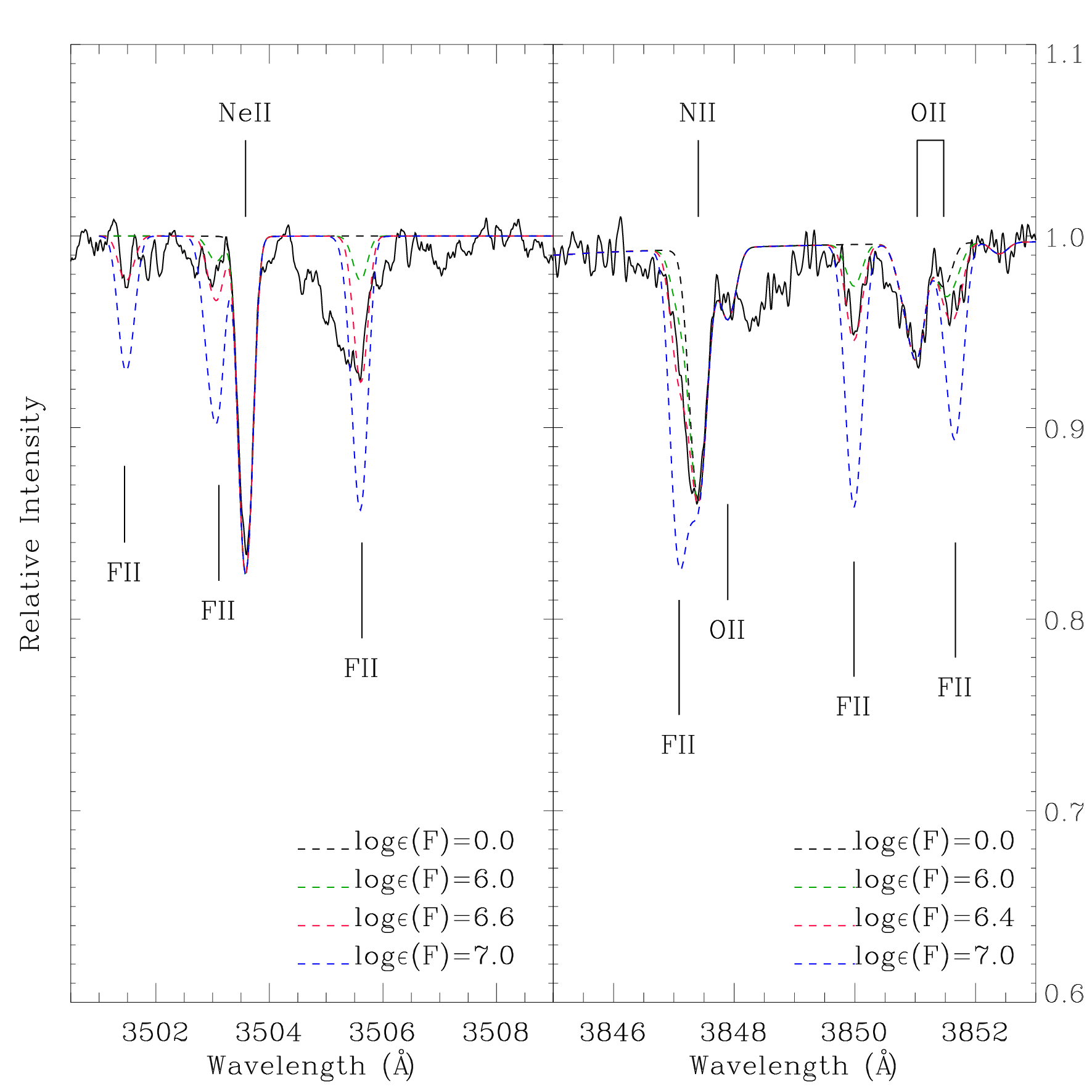}
\caption{\small{ Observed F\,{\sc ii} in 3500\AA \ and 3850\AA \ of LS\,IV+$6^{\degree}$ 2 (solid line) with key lines marked. Synthetic spectra are shown for four fluorine abundanes.}}\label{fig.3}
\end{figure*}

\emph{LS\,IV+$6^{\degree}$ 002.}  The windows at 3505 \AA \ and 3850 \AA \ are both available for
this star.  The F abundance is based primarily on the lines at 3849.986 \AA \ and 3851.7 \AA \ with the two weakest lines at 3501.4  \AA \ and 3503.1 \AA \ providing supporting evidence as to the maximum F abundance 
allowed by these lines (Figure \ref{fig.3}). The 3847.1 \AA \ line in the blue wing of a strong N\,{\sc ii} line appears present but assessment of its strength is dependent on the adopted width of the N\,{\sc ii} line.
The 3505  \AA \ blend of three RMT 3 lines appears to be present at the F abundance provided by other lines but is seriously blended with an unidentified line. The F abundance of $\log\epsilon$(F) = 6.5 seems appropriate for this star.

\begin{figure*}
\center
\includegraphics[scale=0.5]{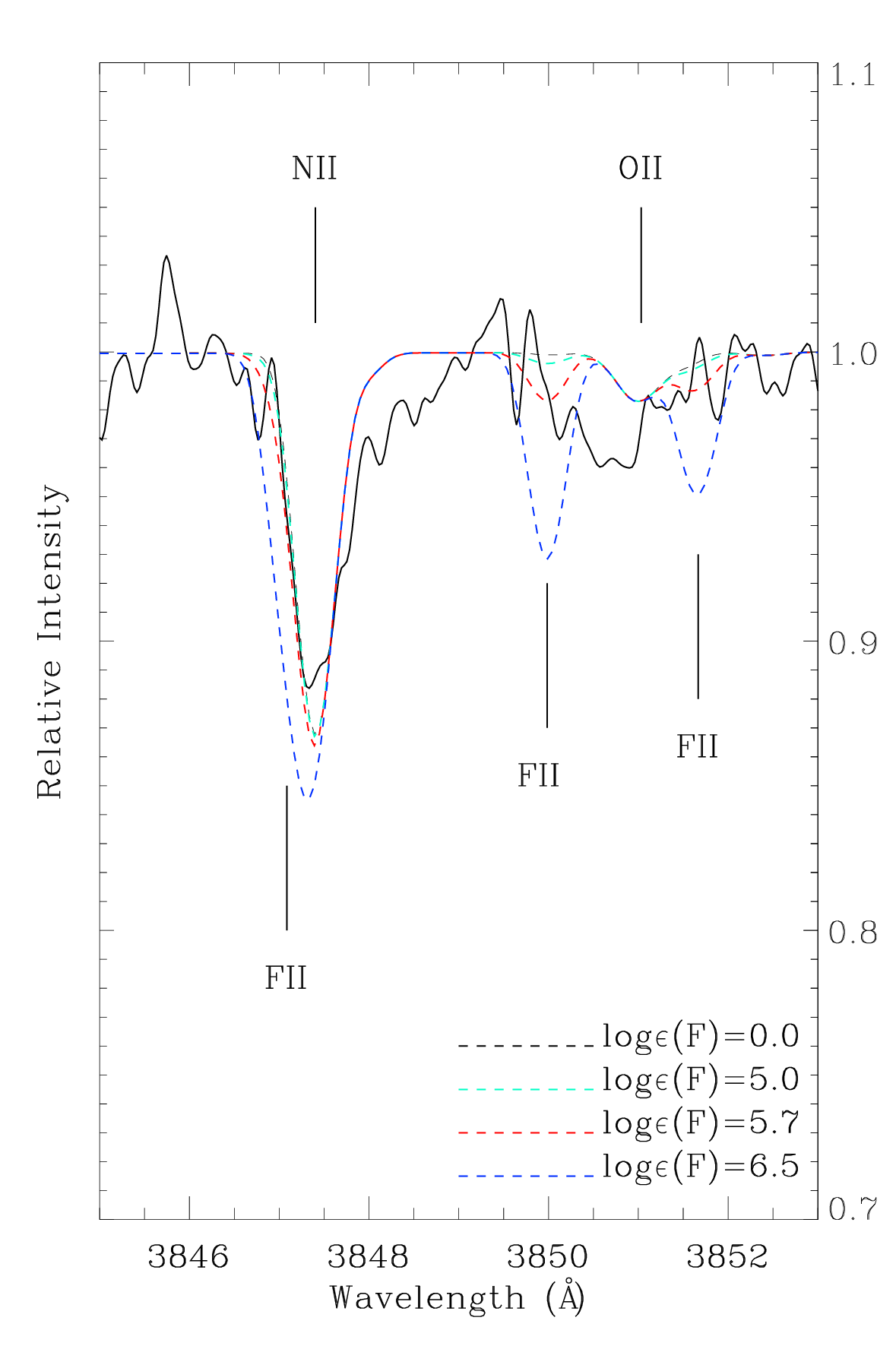}
\caption{\small{Observed F\,{\sc ii} in 3850\AA \ of V652\,Her (solid line) with key lines marked. Synthetic spectra are shown for four fluorine abundanes.}}\label{fig.4}
\end{figure*}

\emph{V652\,Her.}  Only the 3850 \AA \ window is available. Spectrum synthesis does not provide convincing detection of a F\,{\sc ii} line (Figure \ref{fig.4}). An upper limit of $\log\epsilon$(F) = 5.6 is provided by each of the RMT 1 lines.  This star is very clearly F-poor relative to LS IV $+6^o$ 002 (and other F-rich stars).

\begin{figure*}
\center
\includegraphics[scale=0.5]{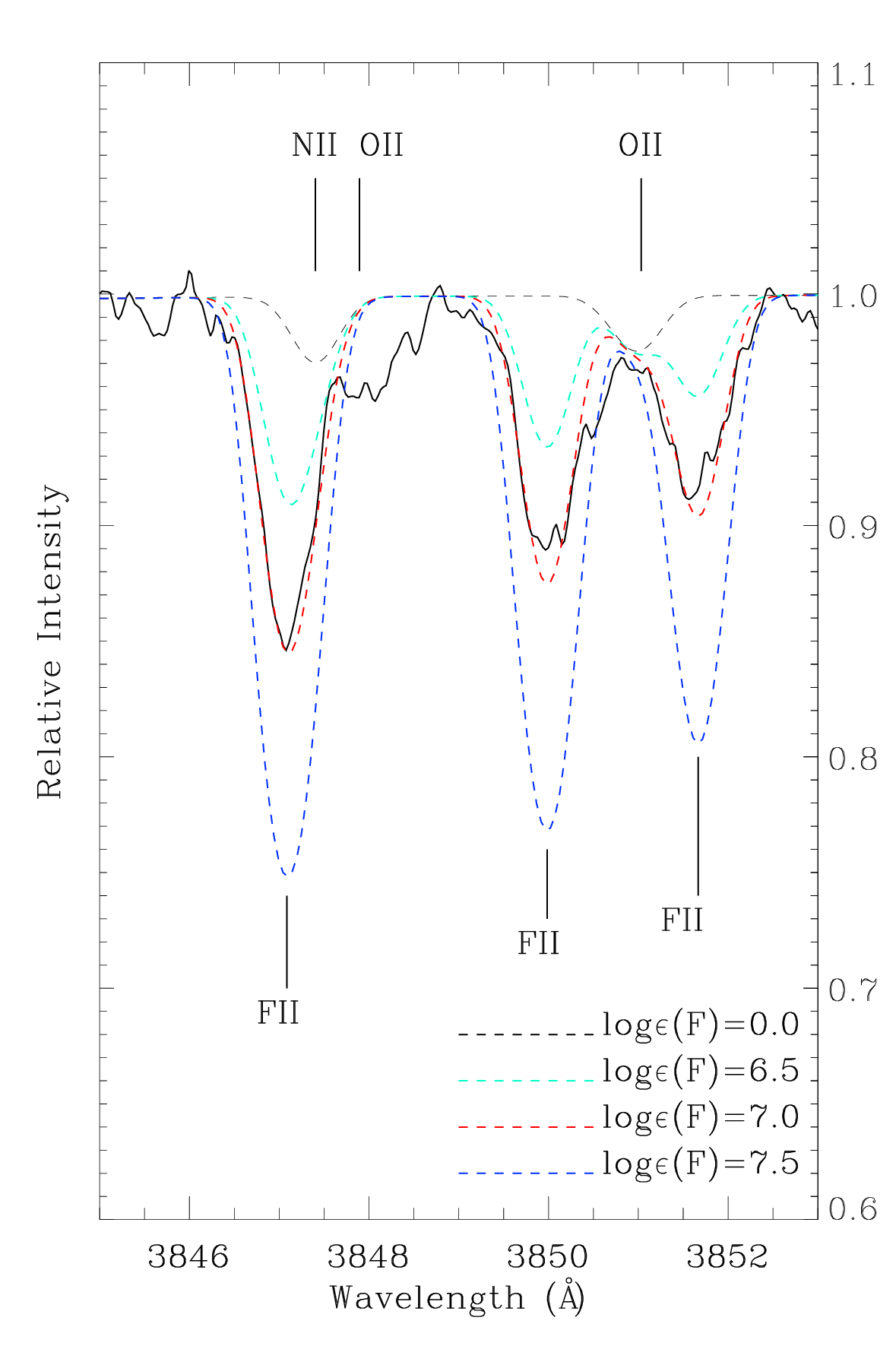}
\caption{\small{Observed F\,{\sc ii} in 3850 \AA \ of V2205\,Oph (solid line) with key lines marked. Synthetic spectra are shown for four fluorine abundanes.}}\label{fig.5}
\end{figure*}

\emph{V2205\,Oph.} The 3505 \AA \ window is not available. In the 3850  \AA \ window, the three
RMT 1 F\,{\sc ii} lines are clearly present with a consistent abundance of $\log\epsilon$(F) = 7.0 (Figure \ref{fig.5}). Blending lines of N\,{\sc ii} and O\,{\sc ii} are pleasingly weak in this star ensuring the consistency of the F abundance from the three lines.

\begin{figure*}
\center
\includegraphics[scale=0.5]{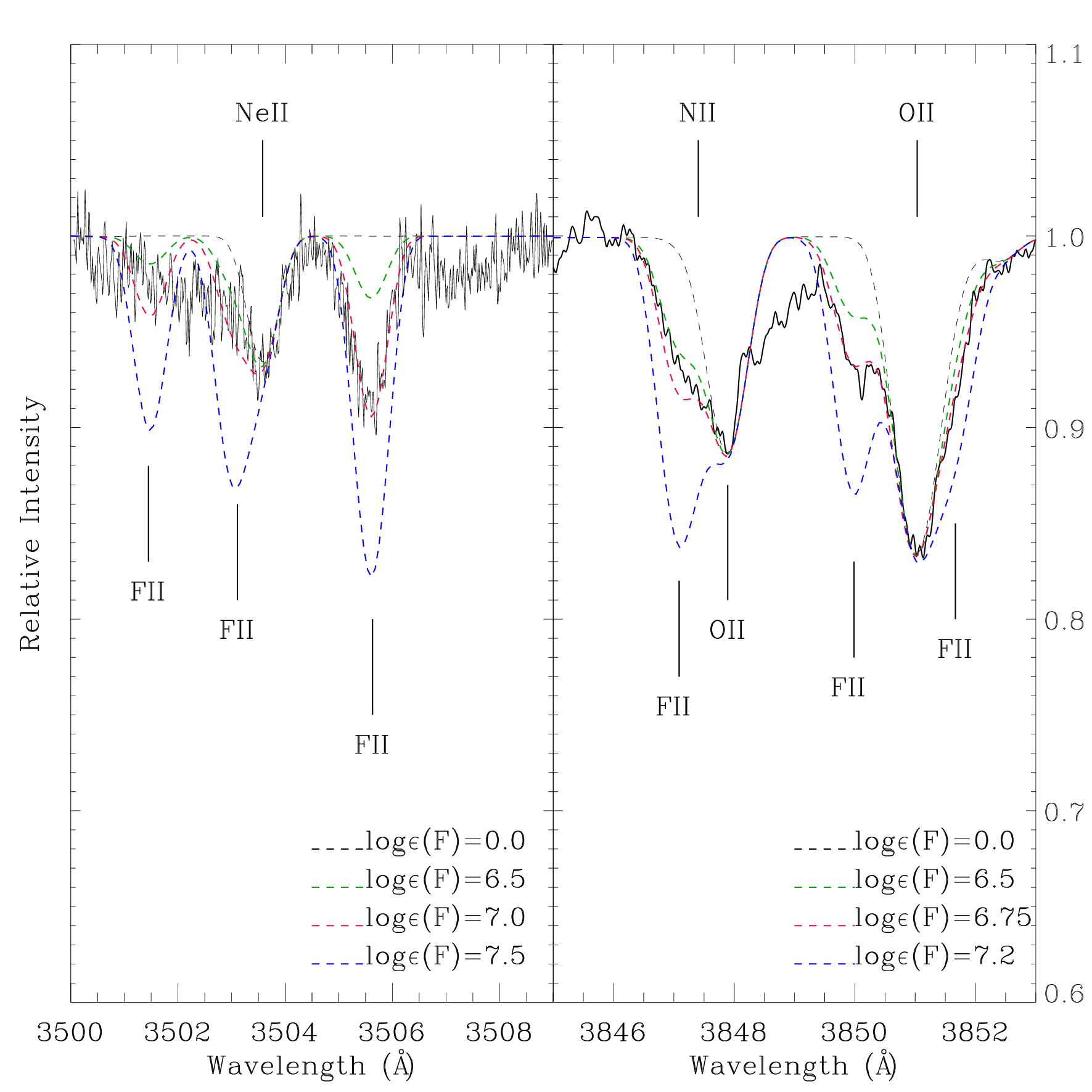}
\caption{\small{ Observed F\,{\sc ii} in 3500\AA \ and 3850\AA \ of DY\,Cen (solid line) with key lines marked. Synthetic spectra are shown for four fluorine abundanes.}}\label{fig.6}
\end{figure*}

\emph{DY\,Cen.} Consistent F abundances are obtained from unblended or relatively unblended lines in both windows (Figure \ref{fig.6}). The unblended 3505.5 \AA \ line provides the F abundance of $\log\epsilon$(F) = 7. The other two lines in the RMT 3 are possibly present and consistent with this abundance. In RMT 1,  the blending by the N\,{\sc ii} and O\,{\sc ii} lines is much stronger than in V2205\,Oph (Figure \ref{fig.5}). The weaker two F\,{\sc ii} lines of this RMT provide a consistent F abundance which is supported by the 3851.7 \AA \ line now seriously blended with the O\,{\sc ii} line. A F abundance of $\log\epsilon$(F) = 6.9 is recommended.

\begin{figure*}
\center
\includegraphics[scale=0.5]{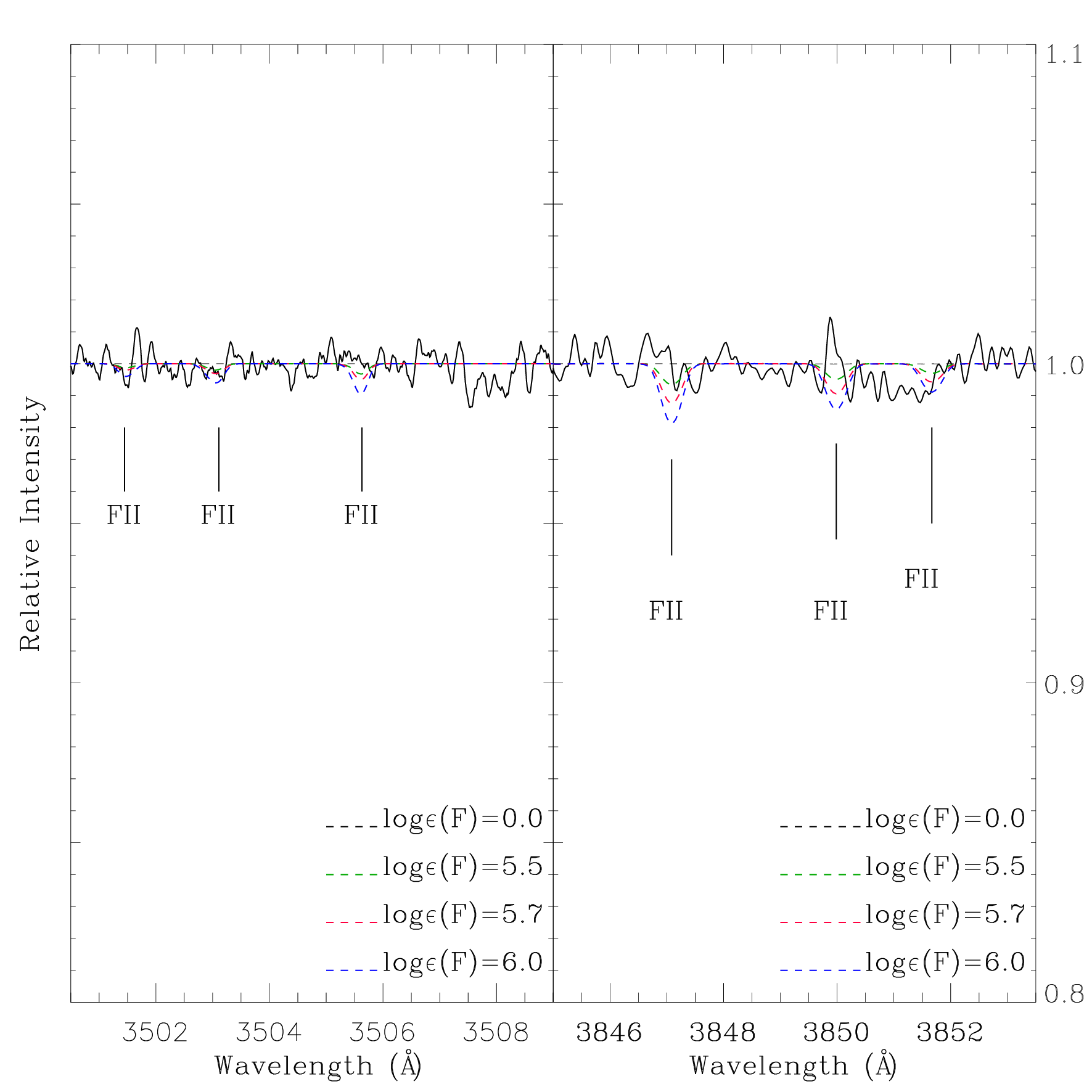}
\caption{\small{ Observed F\,{\sc ii} in 3500\AA \ and 3850\AA \ of HD\,144941 (solid line) with key lines marked. Synthetic spectra are shown for four fluorine abundanes.}}\label{fig.7}
\end{figure*}

\emph{HD\,144941.} The wavelength regions centered at 3505\AA \ and 3850\AA \ are available and do not show detectable F\,{\sc ii} lines in the observed spectrum (see Figure \ref{fig.7}).  An upper limit of $\log\epsilon$(F) = 5.6  is obtained by each of RMT 1 and RMT 3 lines. This star is clearly very F-poor and  similar to the other carbon poor hot EHe star:V652\,Her.

\begin{figure*}
\center
\includegraphics[scale=0.5]{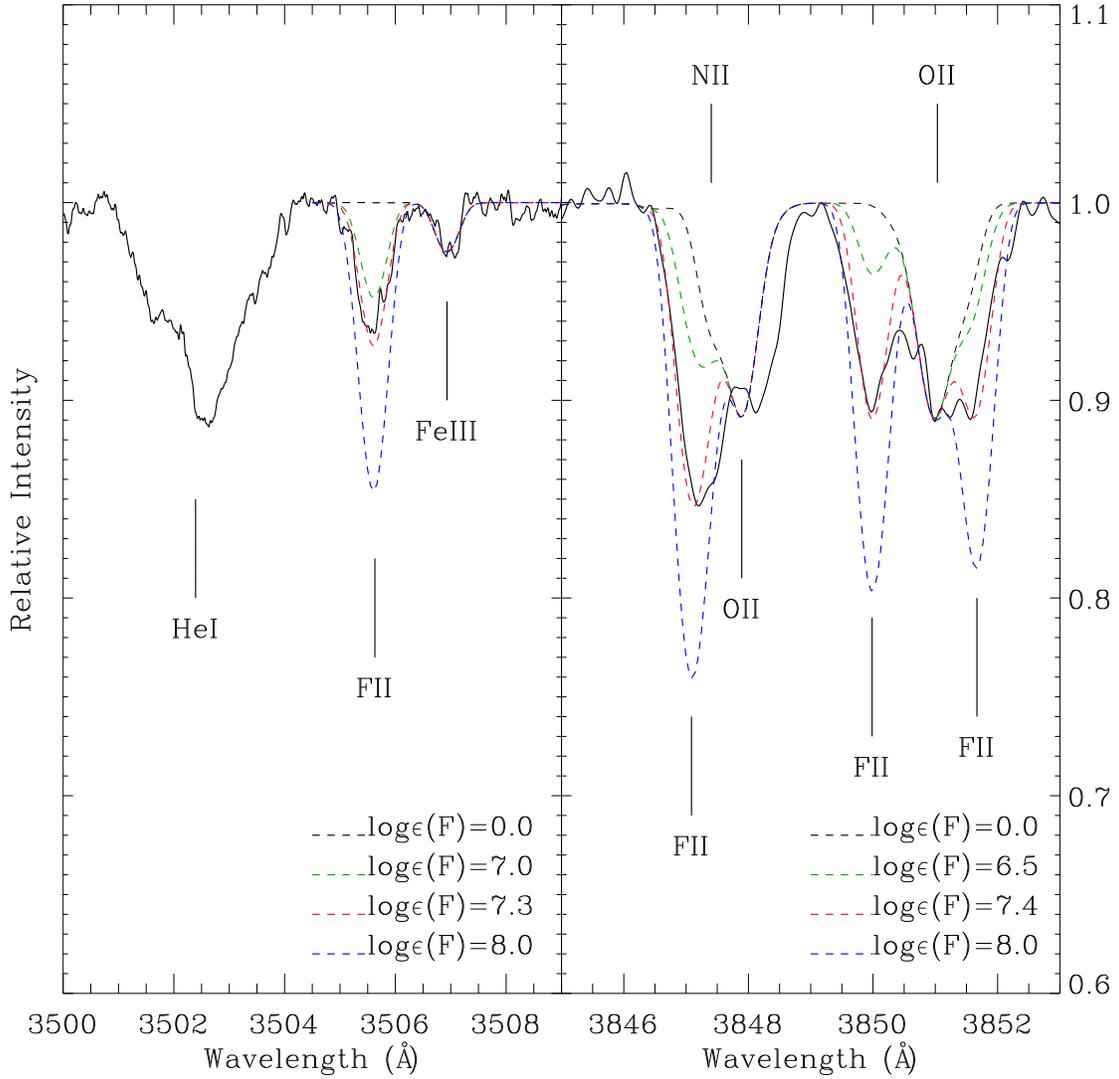}
\caption{\small{ Observed F\,{\sc ii} in 3500\AA \ and 3850\AA \ of LSE\,78 (solid line) with key lines marked. Synthetic spectra are shown for four fluorine abundanes. Note that the He\,{\sc i} line at 3502 \AA \  in the left panel of the above figure is not synthesized due to unavailability of log-$gf$ values in NIST database }}\label{fig.8}
\end{figure*}

\emph{LSE\,78.}  The F abundance is provided by the F\,{\sc ii} lines at 3505.5  \AA \ from RMT 3 and the three lines of RMT 1. The F abundance of $\log\epsilon$(F) = 7.4 reproduces these lines (Figure \ref{fig.8}).

\begin{figure*}
\center
\includegraphics[scale=0.5]{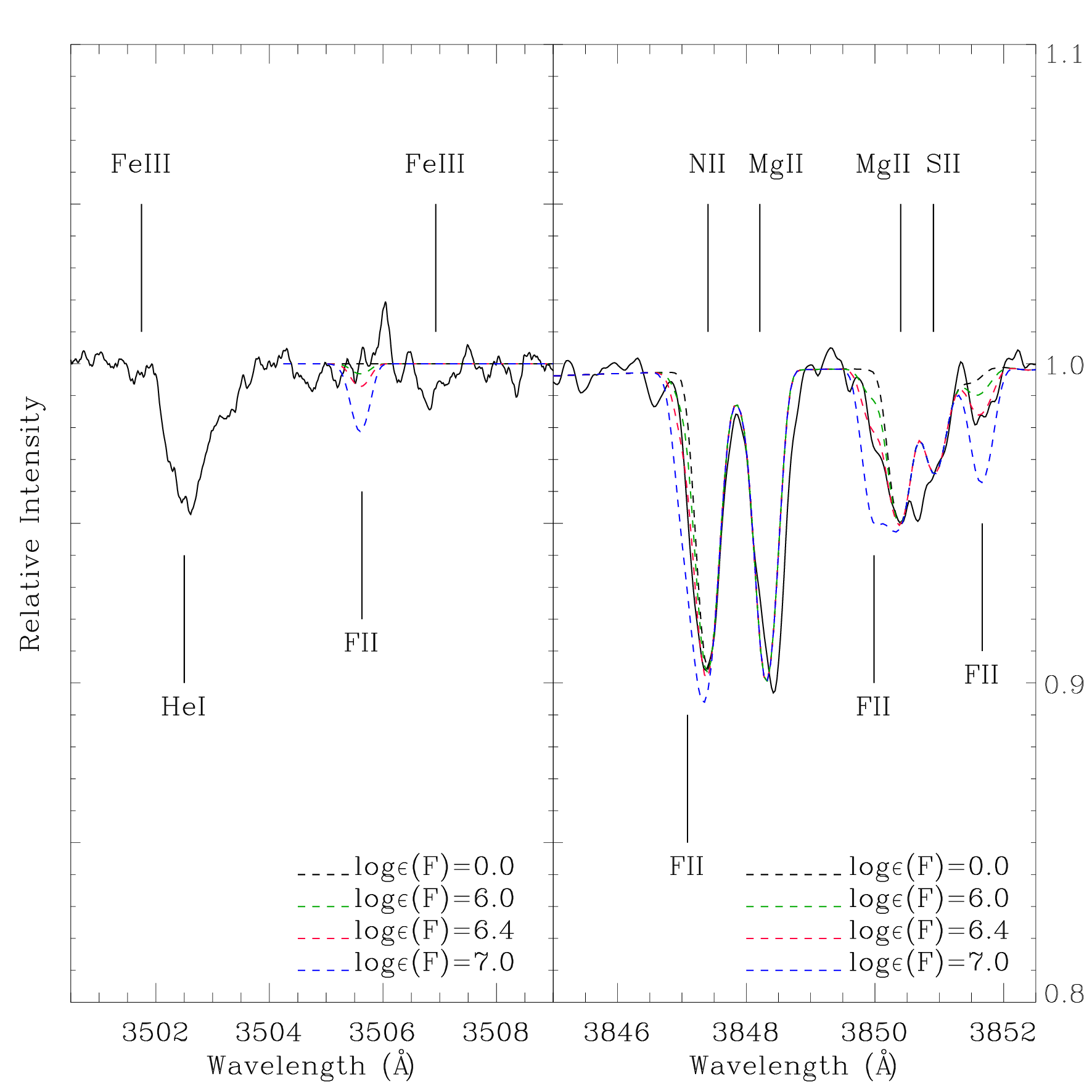}
\caption{\small{ Observed F\,{\sc ii} in 3500\AA \ and 3850\AA \ of BD +$10^{\degree}$\,2179 (solid line) with key lines marked. Synthetic spectra are shown for four fluorine abundanes. }}\label{fig.9}
\end{figure*}

\emph{BD +$10^{\degree}$\,2179.}  In both the 3505 \AA \ and 3850  \AA \ windows, the F\,{\sc ii}
lines are weak (Figure \ref{fig.9}). The 3505.5 \AA \ is not detected and the abundance upper limit for F of $\log\epsilon$(F) = 6.5 is set. In RMT 3, the cleanest line is at 3851.7  \AA \ and gives the F abundance of about $\log\epsilon$(F) = 6.4. The other two lines of this RMT are blended but confirm the abundance of 6.4.

\begin{figure*}
\center
\includegraphics[scale=0.5]{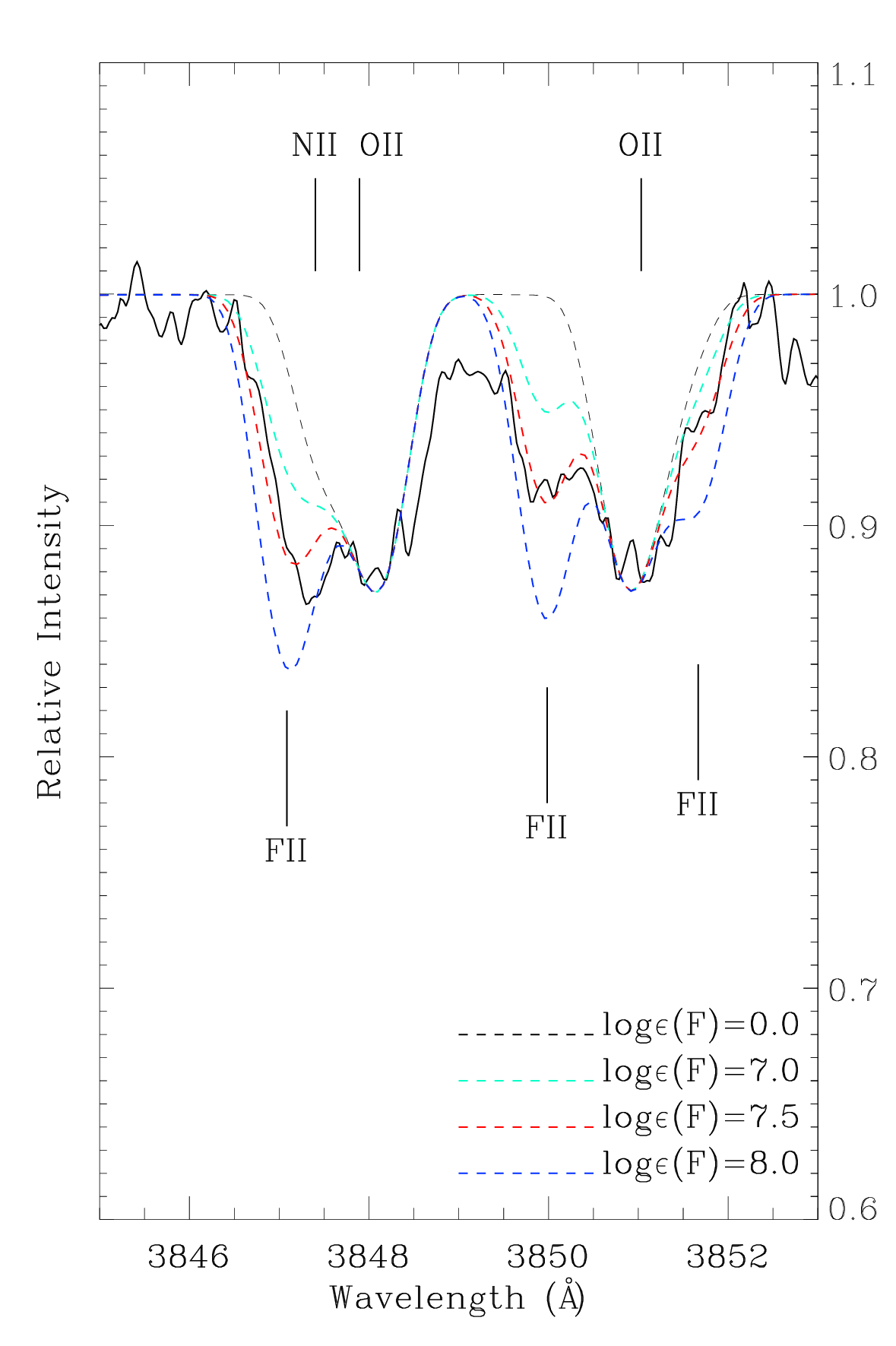}
\caption{\small{Observed F\,{\sc ii} in 3850 \AA \ of V1920\,Cyg (solid line) with key lines marked. Synthetic spectra are shown for four fluorine abundanes. }}\label{fig.10}
\end{figure*}

\emph{V1920\,Cyg.} Only the 3850 \AA \ window is available where the three F\,{\sc ii} lines are blended (Figure \ref{fig.10}). The least blended line at 3850.0 \AA \ gives the F abundance of $\log\epsilon$(F) = 7.5, a value consistent with determinations from the  two more seriuosly blended lines.

\begin{figure*}
\center
\includegraphics[scale=0.5]{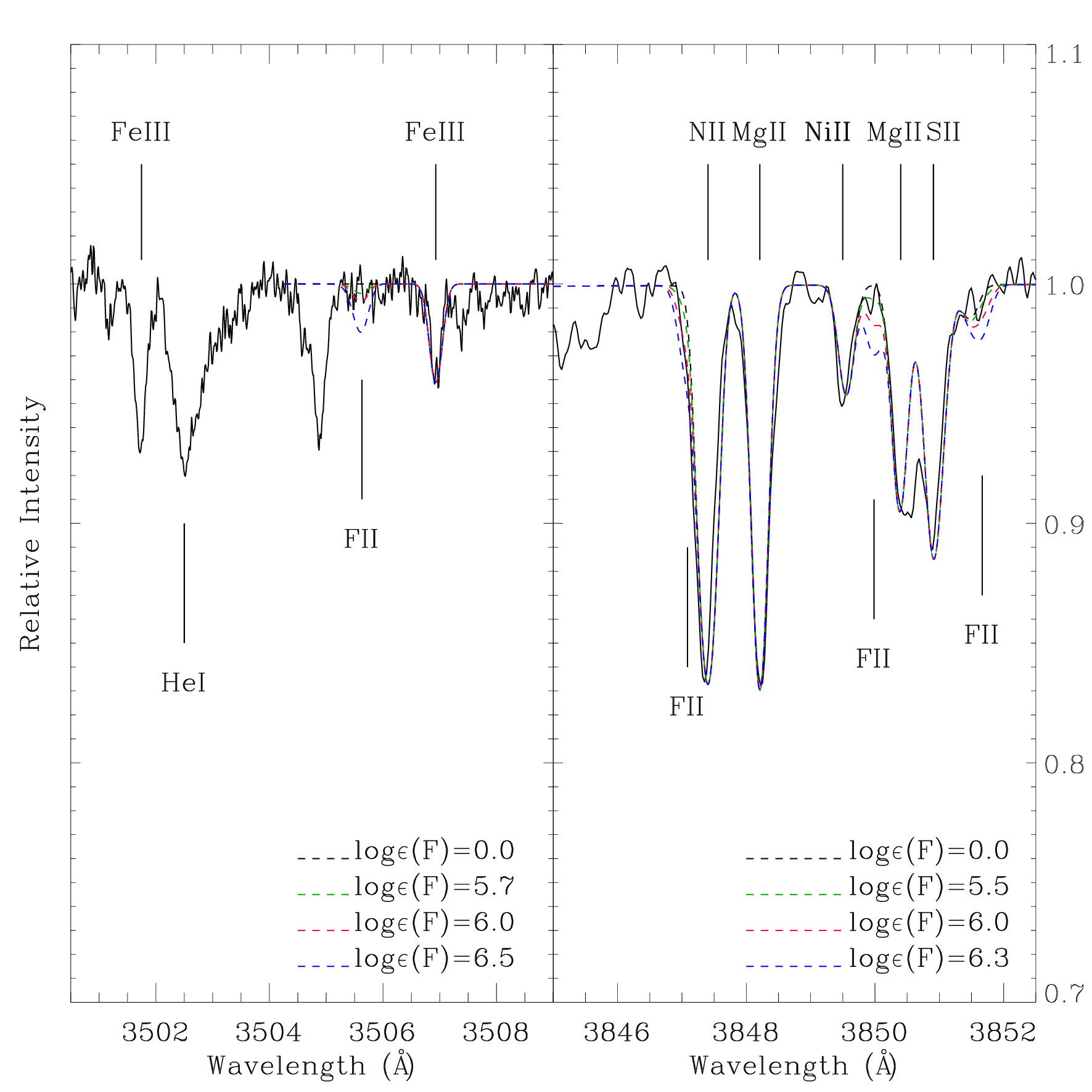}
\caption{\small{ Observed F\,{\sc ii} in 3500\AA \ and 3850\AA \ of HD\,124448 (solid line) with key lines marked. Synthetic spectra are shown for four fluorine abundanes. }}\label{fig.11}
\end{figure*}

\emph{HD\,124448.} Both wavelength regions are available but neither show evidence for the F\,{\sc ii} lines (Figure \ref{fig.11}). The 3505.5 \AA \ is clearly absent.  In the 3850 \AA \ window blends are an issue but the 3850.0  \AA \ and 3851.7 \AA \ lines are absent. An upper limit to the F abundance of $\log\epsilon$(F) = 6.0 may be set.

\begin{figure*}
\center
\includegraphics[scale=0.5]{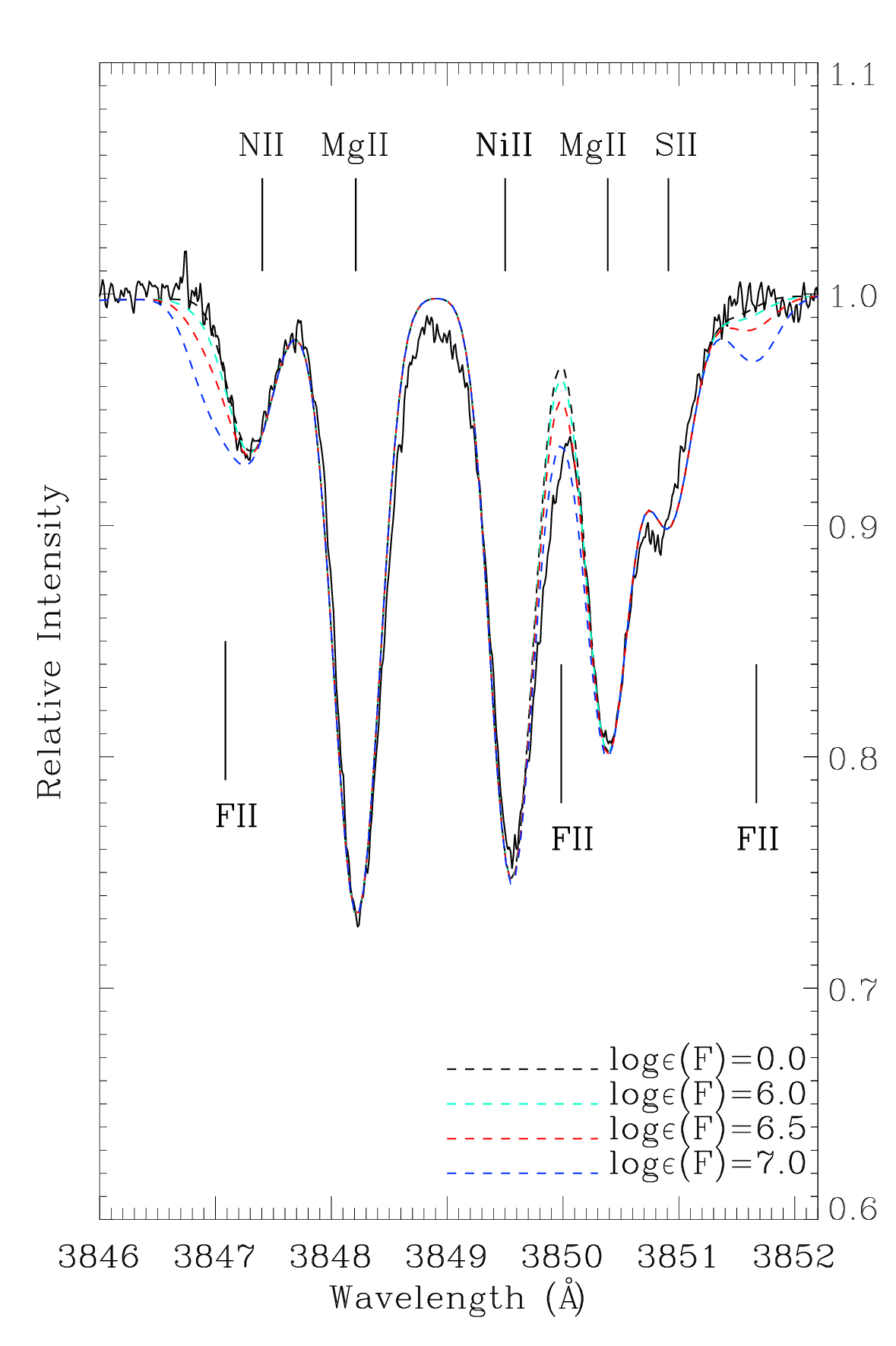}
\caption{\small{ Observed F\,{\sc ii} in 3850 \AA \ of PV\,Tel (solid line) with key lines marked. Synthetic spectra are shown for four fluorine abundanes. }}\label{fig.12}
\end{figure*}

\emph{PV\,Tel.} Only the 3850 \AA \ window is available (Figure \ref{fig.12}). Spectrum synthesis does not provide convincing detection of a F\,{\sc ii} line. An upper limit of $\log\epsilon$(F) = 6.5 is provided by the 3847.1  \AA \ and 3851.7 \AA \ lines.

Our previous investigations of the compositions of  hot EHe stars included analyses of the Ne abundance from Ne\,{\sc i} and/or Ne\,{\sc ii} lines. When both neutral and singly ionized lines were available, it was found that the Ne abundance from the neutral lines was higher than that from the singly-ionized lines: For example, the LTE neon abundances obtained by \citet{Pandey2011} for  V2205\,Oph from Ne\,{\sc i} lines was 0.8 dex higher than obtained from Ne\,{\sc ii} lines. This difference arises from non-LTE effects principally affecting the Ne\,{\sc i} lines, a suggestion thoroughly confirmed by a non-LTE study by  \citet{Pandey2011}. Noting that the atomic structure of the F atom and the detected F\,{\sc i} lines are not dissimilar to the Ne atom and the non-LTE affected Ne\,{\sc i} lines, we attempted to set limits on the non-LTE effects on the F abundances by analysing F\,{\sc i} and F\,{\sc ii} lines in the same star.

\begin{figure*}
\center
\includegraphics[scale=0.5]{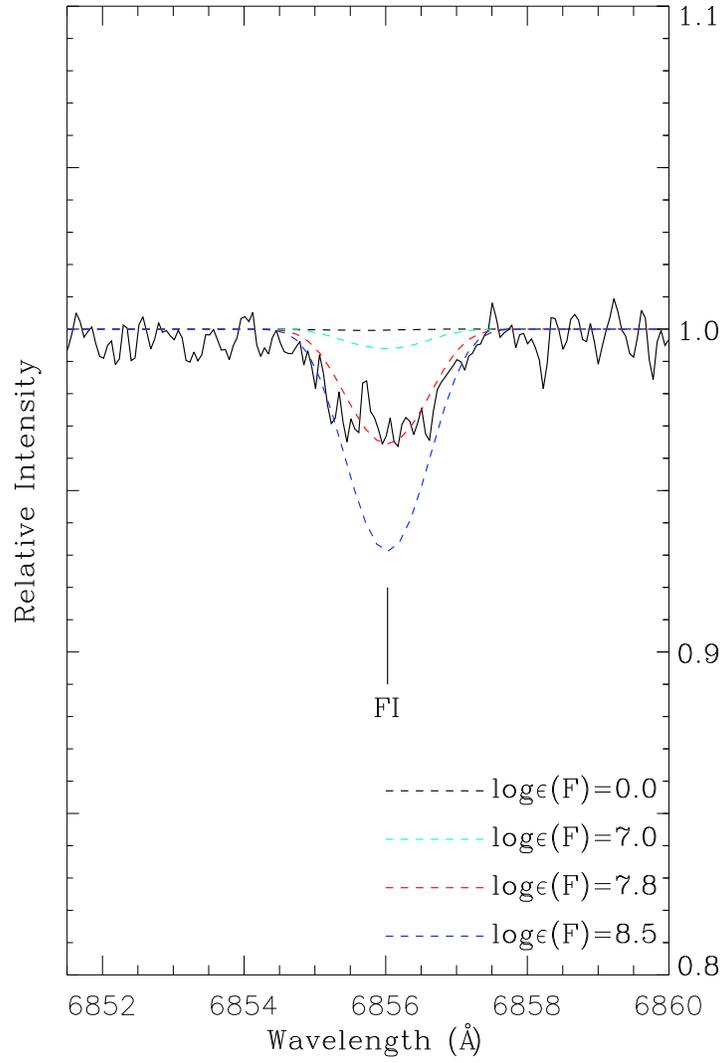}
\caption{\small{ Observed F\,{\sc i} in 6856\AA \ of V1920\,Cyg (solid line) with key lines marked. Synthetic spectra are shown for four fluorine abundances. }}\label{fig.13}
\end{figure*}

\begin{figure*}
\center
\includegraphics[scale=0.5]{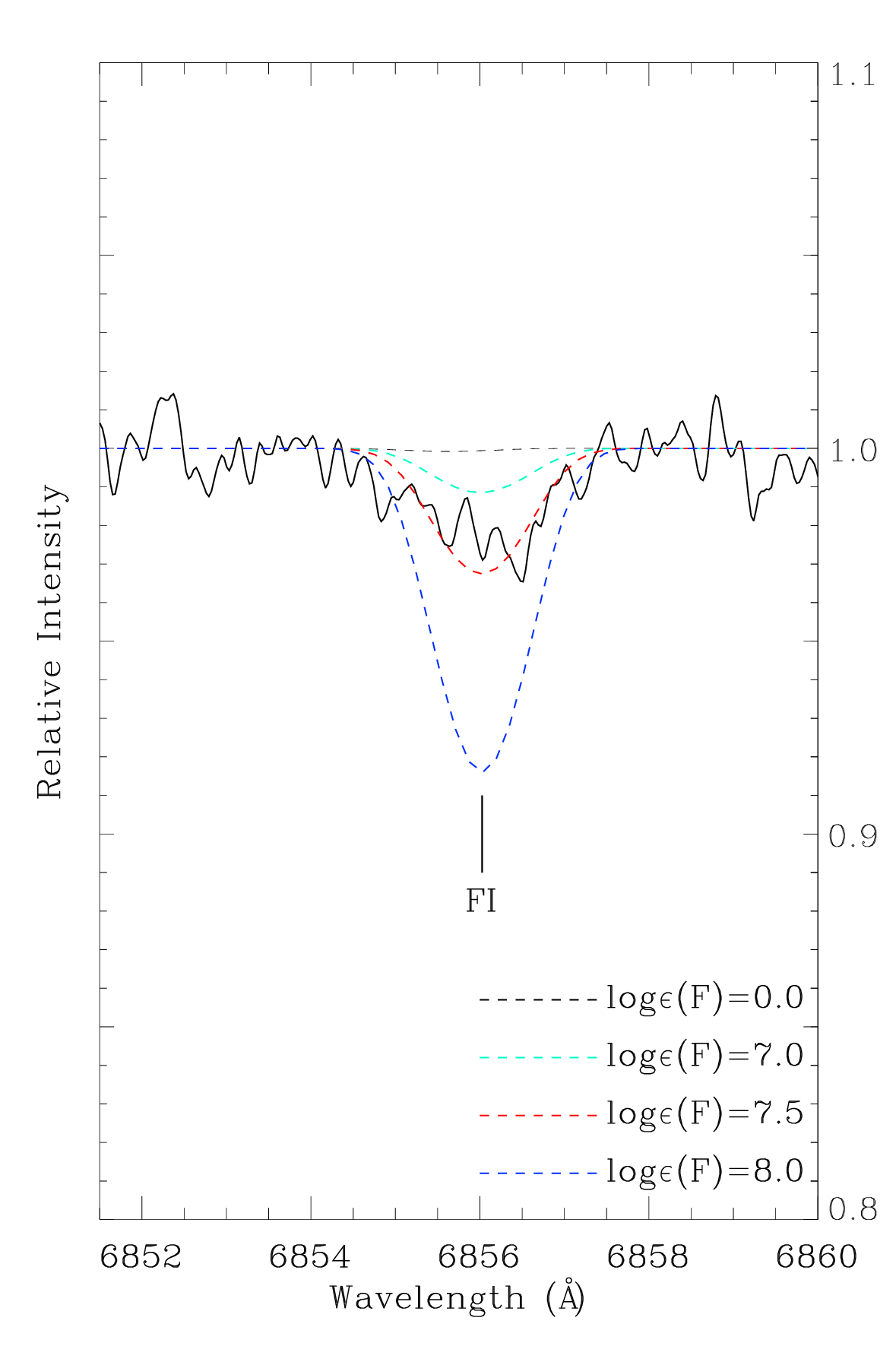}
\caption{\small{ Observed F\,{\sc i} in 6856\AA \ of LSE\,78 (solid line) with key lines marked. Synthetic spectra are shown for four fluorine abundances. }}\label{fig.14}
\end{figure*}

 In the present sample of hot EHes, examination of the spectrum of V1920\,Cyg and LSE\,78 led to the detection of the  F\,{\sc i} at 6856.02 \AA \ line (Figure \ref{fig.13} \& \ref{fig.14} ) with the estimated F abundance of 7.8 and 7.5, respectively. This F\,{\sc i} line at 6856.02 \AA \ is the strongest F\,{\sc i} line and the weaker F\,{\sc i} lines are consistent with it. The abundance difference between that from the F\,{\sc ii} and the F\,{\sc i} lines is $-0.3$ and $-0.1$ dex for V1920\,Cyg and LSE\,78, respectively.These differences limit severely the non-LTE effects in the conditions  prevailing in both V1920\,Cyg and LSE\,78.   As a complementary effort, we have returned to spectra of the cooler EHes where the F abundance is based on the F\,{\sc i} lines to look for F\,{\sc ii} lines. Two stars had effective temperatures sufficiently high with the available spectra possessing adequate S/N in the blue to provide interesting limits on F\,{\sc ii} lines: LSS\,3378 and PV\,Tel. For LSS\,3378, the F\,{\sc ii} lines from multiplet 1 provide the upper limit of 8.0 which is consistent with the determination of 7.3 from the F\,{\sc i} lines, a comparison which provides no information on the non-LTE effects. For PV\,Tel, the F\,{\sc ii} lines set the upper limit 6.5 which  is a significant improvement over the limit ($\le 7.2$) reported by \citet{Pandey2006f} from the F\,{\sc i} lines . Other EHes where - in the future one might get both F\,{\sc i} and F\,{\sc ii} lines are LS IV -$1^{\degree}$\,002 and LSS\,4357. A key point to note is that the level of the F overabundances in the EHes is around a factor of 1000 but the non-LTE effects, if comparable to those on the Ne\,{\sc i} lines, are less than a factor of 10. Thus, a very significant overabundance of F in EHes and RCBs is not in doubt because of the present lack of non-LTE calculations for the fluorine atom.

\subsection{Fluorine enrichment}

Fluorine abundances across the sample of ten hot EHes which range from 7.5 to an upper limit of 5.6 are shown in Figure \ref{fig.15} as a function of Fe abundance where the two C-poor stars (V652\,Her and HD\,144941) are distinguished. The F overabundance is remarkable. Fluorine abundances  for the cool EHes \citep{Pandey2006f} and the majority and minority RCBs \citep{Pandey2008,Hema2017} are added to Figure \ref{fig.15}.  The spread of F abundances across the total sample of H-deficient stars far exceeds the errors of measurement.  Figure \ref{fig.15} suggests that a few stars may have a F abundance much lower than the typical EHe and RCB. This minority includes the two C-poor hot EHe stars V652\,Her and HD\,144941, the minority RCB V854\,Cen with $\log\epsilon$(Fe)=5, the majority RCB XX\,Cam and possibly also the hot EHe HD\,124448.

 In Figure \ref{fig.15}, the stars' initial F abundance is assumed to correspond to [F/Fe] $=$ 0 over the range $\log\epsilon$(Fe) from about 7.5 to 5.0 with the trend starting with a solar F abundance of  $\log\epsilon$(F) $= 4.4$ at $\log\epsilon$(Fe) = 7.5 (see below). A star's Fe abundance is assumed to be its initial Fe abundance as a H-normal star. Relative to the assumed F-Fe trend, the typical F overabundance at the solar Fe abundance is 500 and this increases to nearly 2000 for the most metal-poor of the H-deficient stars, namely the minority RCBs and the cool EHe  FQ\,Aqr. This is an extraordinary overabundance for any element in any star! The common F overabundance among EHe and RCB stars indicates that, as long suspected, these H-deficient stars are probably closely related. (Fluorine abundances have  not been measured for HdC stars.) In sharp contrast, the H-deficient spectroscopic binary KS Per has an upper limit to the F abundance consistent with its initial abundance \citep{Pandey2006f} confirming expectations that massive hot binaries like KS Per have an entirely different evolutionary history to the EHes and their relatives.

Initial abundances for interpretations of F abundances in EHe and RCB stars are based on the solar abundance and abundances in red giants in the Galactic disk. Solar determinations of the F abundance are determined from infrared lines of HF in sunspot spectra: \citet{Maiorca2014} report a solar F abundance of $\log\epsilon$(F) = 4.40$\pm0.25$ in fine agreement with the abundance of 4.42$\pm0.06$ obtained from meteorites \citep{Lodders2009}. F abundance measurements from HF lines in spectra of red giants in the Galaxy suggest that [F/Fe] $\simeq 0.0$ over the [Fe/H] interval  $0$ to $-1.5$ spanning  the interval covered by the EHe and RCB stars \citep{Nault2013,Li2013,Maiorca2014,Jonsson2014,Jonsson2017,Guerco2019}. Current uncertainty in the F vs Fe relation in Galactic normal stars should not affect the comparison of compositions of EHe, RCB and HdC stars with theoretical evolutionary scenarios; the F overabundance, in particular, dwarfs the current uncertainty.

\begin{table*}[ht]
\caption{Elemental abundances of hot EHes.}
 \begin{center}
\begin{threeparttable}
\resizebox{\textwidth}{!}{%
\begin{tabular}{lrrrrrrrl}\hline\label{Table.5}
Star name & \multicolumn{7}{c}{log$\epsilon$(X)} & Ref\\
\cmidrule{2-8}

                          &   C   &   N  &   O   &  Ne  &  F  &	 Fe  &  Zr  & \\\midrule
          																
LS\,IV+$6^{\degree}$ 2    &	 9.4  &	 8.3 &	8.2	 &  8.7 &  6.5 	&	7.1	  &	  \nodata 	  &	 P11\tnote{1}\\  
          																
V652\,Her           	  &	 7.0  &	 8.7 &	7.6	 &	8.1	&  $\leq$ 5.6 &	  7.1  	  &	  \nodata 	  &	 P17\tnote{2} \\            
          																
V2205\,Oph                &	 9.1  &	 7.8 &	8.0	 &	8.2 &  7.0	  &	6.6	  &	      \nodata 	  &	 P11\tnote{1 }\\    
          																
DY\,Cen                   &	 9.6  &	 7.8 &	9.0	 &	8.0 &  6.9	  &	6.0	  &	      \nodata 	  &	 P14\tnote{3} \\   
   																
HD\,144941                &	 6.9  &	 6.4 &	7.1	 &  7.2 & $\leq$ 5.6 & $\leq$ 6.6 &	\nodata & P17\tnote{2} \\ 
																
LSE\,78                   &	 9.4  &	 8.3 &	9.4	 &	8.7 &  7.4	     &	6.8	      &	3.5	    & P11\tnote{1} ; P06a\tnote{4}\\
																
BD +$10^{\degree}$\,2179    &	 9.3  &	 8.1 &	7.9	 &  7.9 &  6.4	     &	6.2	      &	 $\leq$  2.6  &  P11\tnote{1} ; P06a\tnote{4} \\
																
V1920\,Cyg                &	9.6	  &	 8.6 &	9.9	 &	8.5 &  7.5	     &	6.8	      &	3.7	     &	 P11\tnote{1} ; P06a\tnote{4} \\
																
HD\,124448                &	9.1	  &	 8.7 &	8.3	 &  7.7 &  $\leq$ 6.0	    & 	7.2  	  &	2.7	  &	 P11\tnote{1} ; P06a\tnote{4} \\ 

PV\,Tel                   & 9.2   &  8.6 &  8.8  &  7.6 &  $\leq$ 6.5   &   7.0      & 3.1 &   P11\tnote{1} ; P06a\tnote{4} \\ \hline

\end{tabular}}
\begin{tablenotes}
\item [1] \citep{Pandey2011}
\item [2] \citep{Pandey2017}
\item [3] \citep{Pandey2014}
\item [4] \citep{Pandey2006}
\end{tablenotes}
\end{threeparttable}
\end{center}
\end{table*}

\subsection{Fluorine and other elements}

\begin{figure*}
\center
\includegraphics[scale=0.7]{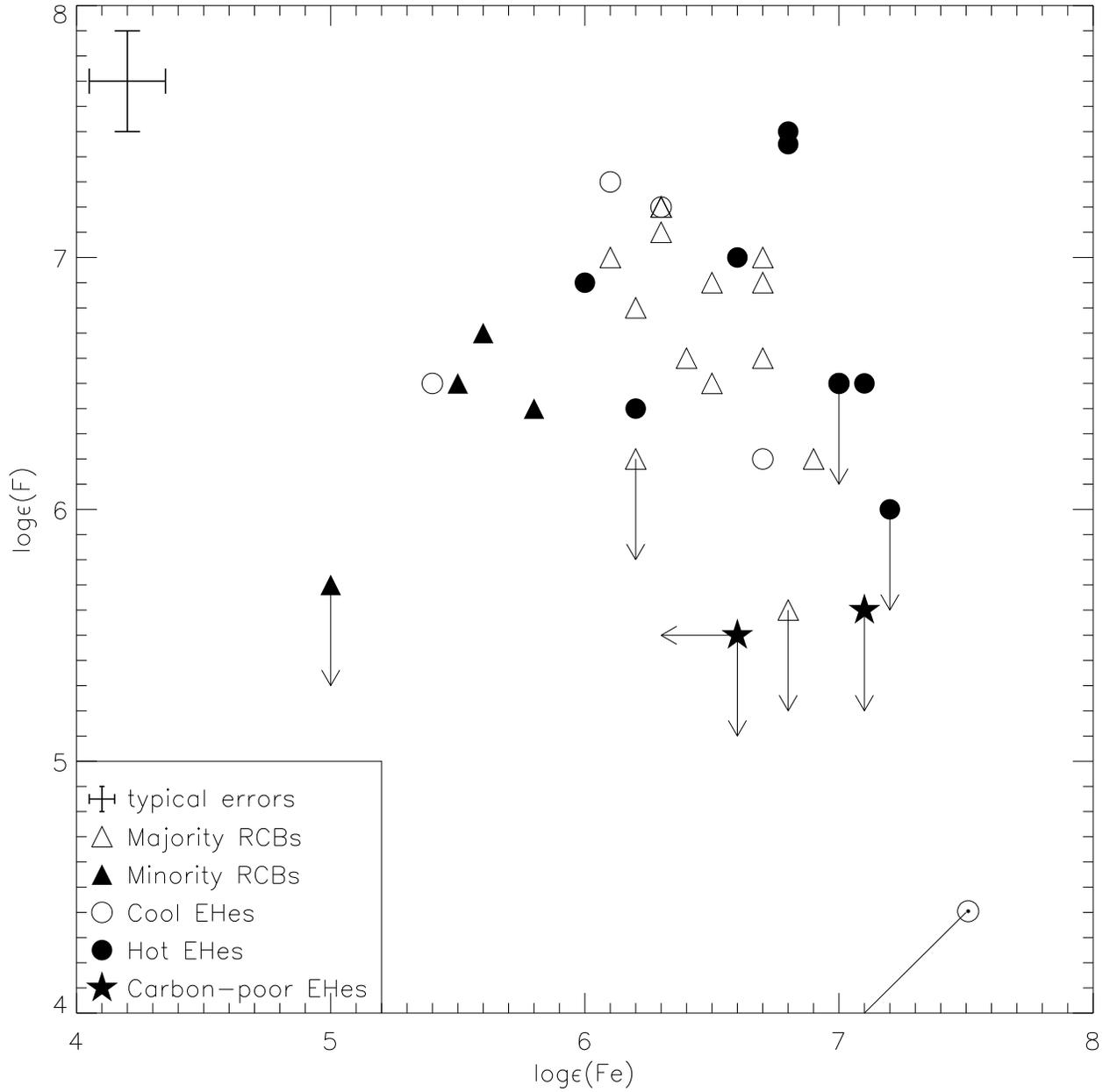}
\caption{\small{log$\epsilon$(F) versus log$\epsilon$(Fe) for hot EHes, cool EHes, and RCBs. The symbols representing different group of stars are showed. The encircled dot symbol represents the sun and the solid line represents locus of the solar  F/Fe ratio. }}\label{fig.15}
\end{figure*}

In searching for an explanation for the hot EHes, the cool EHes and the RCBs, it is helpful to identify relationships, if any, between the abundances of key elements.  No modern  analysis for elemental abundances is available for the HdC stars whose spectra is dominated by molecular bands. Consideration of the elemental abundances for RCB stars must recognize that the available analyses of \citep{Asplund2000} identified `the carbon problem'. Opacity in the atmosphere of a  RCB star appears to be dominated by continuous absorption from excited levels of the neutral carbon atom. Since the many absorption lines of the neutral carbon atom also  arise from excited levels, the predicted strength of  weak C\,{\sc i} lines is almost independent of the principal atmospheric parameters, that is effective temperature, surface gravity and the C/He ratio: however, the predicted equivalent widths of weak C\,{\sc i} lines is a factor of 0.6 dex stronger than observed. This discrepancy defines the carbon problem, e.g., model atmospheres computed for a C abundance of 9.5 (equivalent to a C/He ratio of 1\%) return a C abundance of   8.9 from weak C\,{\sc i} lines, The carbon  problem's  implications for abundances and abundance ratios are discussed but not resolved in an extensive investigation of possible solutions by \citet{Asplund2000}.  Although some proposed resolutions of the carbon problem should have minimal effect on elemental abundances and particularly on abundance ratios, abundances for RCB stars  should be used with reservation in effecting comparisons with compositions of EHe stars. EHe stars are not subject to a carbon problem.  (Abundances for RCBs are used in Figure \ref{fig.15} where it is clear that the F and Fe abundances of EHe and RCB stars provide overlapping distributions; the Fe and F abundances for RCB stars are not both overestimated  by 0.6 dex.)

\begin{figure*}
\center
\includegraphics[scale=0.6]{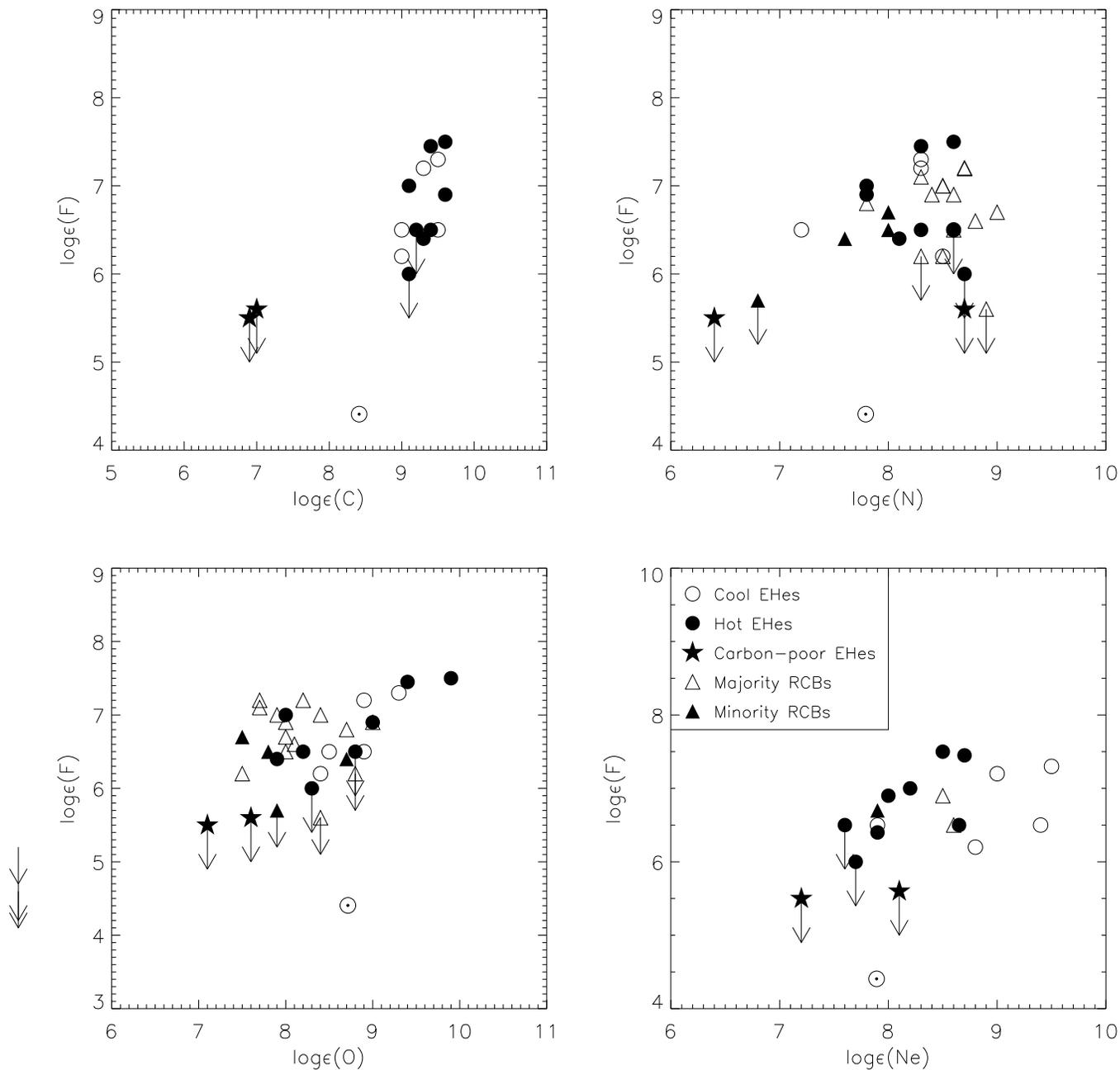}
\caption{\small{Observed log$\epsilon$(F) versus log$\epsilon$(X) for EHes and RCBs from where X = C, N ,O and Ne respectively.  The encircled dot symbol represents the sun.}}\label{fig.16}
\end{figure*}

The likely relationship between F abundances and abundances of C, N, O and Ne  in EHes are shown in Figure \ref{fig.16}.  Abundances for the RCB stars generally confirm results for the EHes.  In the case of C, the spectroscopic C abundances (primarily from \citet{Asplund2000}) are systematically 0.6 dex in the mean  lower than for the EHes  because of the carbon problem.  For N, the N abundances from \citet{Asplund2000} and \citet{Hema2017}, and the F abundances from \citet{Pandey2008} and \citet{Hema2017} for the RCBs, overlap well with the abundance spread provided by the EHes. For the EHes, the O-F relation may suggest a positive correlation with the RCB stars possibly superimposed on this correlation but lacking stars with the extreme O  ($>9.1$) abundances. Neon abundances are available from LTE analysis of Ne\,{\sc i} lines for  four RCBs - Y\,Mus and V3795\,Sgr reported by \citet{Asplund2000}; V532\,Oph and ASAS$-$RCB$-$8 by \citet{Hema2017}- but the F abundance has been reported only for three : V3795\,Sgr\citep{Pandey2008}, and  V532\,Oph and ASAS$-$RCB$-$8 \citep{Hema2017}. The minority RCB V3795\,Sgr with the reported Ne abundance falls amongst the  (Ne,F) abundances for  hot EHes but an anticipated non-LTE reduction of about 0.7 dex  to the Ne abundance would suggest V3795 Sgr is Ne-poor for its F abundance. Whereas for the two majority RCBs V532\,Oph and ASAS$-$RCB$-$8, the non-LTE reduction of 0.7 dex, directly places them in (Ne,F) distribution of hot EHes. LTE neon abundances from Ne\,{\sc i} lines are  available for five cool EHes  \citep{Pandey2001,Pandey2006r} and are compared with fluorine in Figure \ref{fig.16}. Except for the cool EHe FQ\,Aqr, the neon abundances w.r.t fluorine for the cool EHes appear systematically higher than (Ne,F) abundances traced by the hot EHes. Clearly, the anticipated non-LTE correction of 0.7 dex will place them with the hot EHes. Also the same non-LTE correction on FQ\,Aqr would suggest that it is Ne-poor for its F-abundance just like the minority RCB V3795\,Sgr.

\begin{figure*}
\center
\includegraphics[scale=0.45]{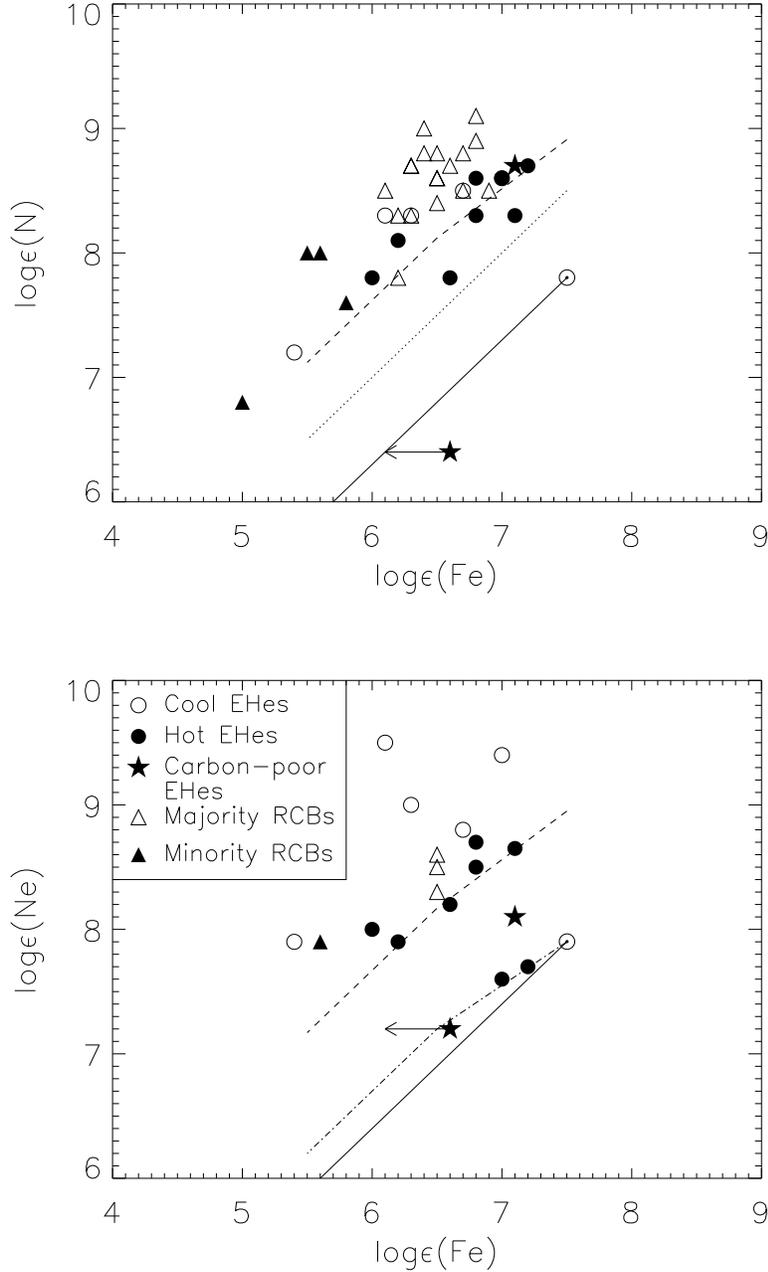}
\caption{\small{Observed log$\epsilon$(N) and log$\epsilon$(Ne) versus log$\epsilon$(Fe) for EHes and RCBs. The encircled dot symbol in each plot corresponds to solar value with the solid line giving the locus of solar ratio N/Fe ad Ne/Fe respectively . The dashed line in the plot of log$\epsilon$(N) vs log$\epsilon$(Fe) is the predicted nitrogen  after full conversion of initial C, N and O to nitrogen in CNO cycle where initial O  is determined from the relation of [$\alpha$/Fe] vs [Fe/H] for normal disk and halo stars given by \citet{Ryde2004}. The dotted line in the same plot is the predicted nitrogen due to conversion of inital C and N to nitrogen in the CN cycle. In the plot of log$\epsilon$(Ne) vs log$\epsilon$(Fe) the dot-dashed line gives the locus of initial neon values taken from the relation of [$\alpha$/Fe] vs [Fe/H] for normal disk and halo stars \citep{Ryde2004}. In the same figure the dashed line is the locus  giving of sum of initial C, N,O and Ne.}}\label{fig.17}
\end{figure*}

Independently of the F abundances,  relations between the N-Ne-Fe abundances provide clues to the stars' nucleosynthetic history (Figure \ref{fig.17}).  H-burning by the CN-cycle increases the N abundance at the expense of C and the ON-cycle provides additional N at the expense of O. In predicting the N abundance from CNO-cycling, initial C and  N abundances are assumed to follow the relation [C/Fe]  = [N/Fe] = 0. Initial O abundances are taken from Ryde \& Lambert’s  (2004) [$\alpha$/Fe] vs [Fe/H]  relation for disk stars with O treated as a typical $\alpha$-element. Nitrogen is supposed here to be the dominant product of CNO-cycling.

Nitrogen abundances as a function of Fe abundances are shown in Figure \ref{fig.17} for the cool  and hot EHes against three possible relations: (i) the initial N vs Fe relation, (ii) the N abundance vs Fe relation expected if the N abundances arises from the sum of the initial C and N abundances, (iii) the N abundances resulting from the sum of the initial C, N and O abundances.  With the clear exception of the C-poor hot EHe HD\,144941,  the N and Fe abundances are distributed along  line (iii) indicating that N is a product of severe CNO-cycling in a H-rich region.  Two hot EHe stars appear closer to the CN-cycling than to the CNO- cycling prediction. With the single exception of the C-poor HD\,144941, the atmosphere of a hot EHe star appears severely contaminated with material exposed to CNO-cycling or possibly in two stars to CN-cycling.

LTE nitrogen abundances for cool EHes \citep{Pandey2001,Pandey2006r} obtained from both N\,{\sc i} and N\,{\sc ii} lines track the N vs Fe trend defined by the majority of the hot EHes from N\,{\sc ii} lines corrected for non-LTE effects. Nitrogen abundances for RCB stars \citep{Asplund2000,Hema2017} from N\,{\sc i} lines but not corrected for non-LTE effects provide N abundances higher than those in the cool and hot EHe stars. This offset arises partially from the lack of a correction for non-LTE effects for the RCB stars and mainly be a symptom related to the carbon problem. Correction for non-LTE effects may lower the RCBs N abundances. In summary, the majority of the H-deficient stars in the RCB-EHe sequence have a N abundance indicative of severe CNO-cyling with the N abundance equalling the initial sum of the C, N and O abundances for a star's Fe abundance.

Neon is severely overabundant in  EHes: Figure \ref{fig.17} (bottom panel) shows the Ne abundances for the hot EHes \citep{Pandey2011,Pandey2014,Pandey2017} where results come from Ne\,{\sc i} lines in all stars and Ne\,{\sc ii} lines in the few hottest stars. The abundance analysis included non-LTE effects which were substantial for Ne\,{\sc i} lines but small for Ne\,{\sc ii} lines.  When lines from the neutral atom and the singly-charged ion were both available, the Ne abundance estimates after non-LTE corrections were in good agreement: see Table 2 of \citet{Pandey2011}.

As discussed earlier neon abundances from Ne\,{\sc i} lines are available for five cool EHe \citep{Pandey2001,Pandey2006r} and four RCBs \citep{Asplund2000,Hema2017}. Ne abundances for the cool EHes are clearly systematically higher than for the hot EHes, for example,  LS IV $-14^o $\, 109 has a Ne abundance of 9.4 for a Fe abundance of 6.9 (Figure \ref{fig.17}).  Such a systematic offset from abundances for the hot EHes is likely due to neglect of the  non-LTE effects on the Ne\,{\sc i} lines.  For the RCBs, Ne with its LTE abundance from Ne\,{\sc i} lines, falls in Figure \ref{fig.17} slightly above the upper boundary of the points from the hot EHes. Application of the non-LTE corrections should place the RCBs and cool EHes with the hot EHes. Then, a majority of hot and cool EHes   and the RCBs have a Ne  abundance  corresponding closely to the sum of the initial C + N + O + Ne abundances.

The upper bound for the Ne vs Fe relation is here set by the condition that the Ne abundance is the sum of the initial C + N + O + Ne abundances which differs only slightly from the C + N + O sum used in the N vs Fe panel. Initial Ne abundances are again taken from \citet{Ryde2004}'s [$\alpha$/Fe] vs [Fe/H] relation for disk and halo stars. Identification of Ne abundances with this sum implies that material has been exposed to temperatures beyond those generally required for H-burning and the product $^{14}$N has been processed by successive $\alpha$-captures to $^{22}$Ne seemingly with near 100\% efficiency. The majority of stars in the (Ne,Fe) panel fall along the (C+N+O+Ne) limit. Among the hot EHes, HD\,124448 and PV\,Tel and possibly also the C-poor and Fe-poor HD\,144941 display a Ne abundance consistent with the star's initial abundance. HD\,124448 and PV\,Tel have N abundances indicating conversion of initial C + N + O to N by H-burning but both appear to have avoided production of Ne by $\alpha$-captures. For V652\,Her, the other C-poor hot EHe, the lower Ne abundance implies either less than complete burning of the $^{14}$N from CNO-cycling to $^{22}$Ne and/or partial destruction of the $^{22}$Ne by $\alpha$-captures. Note that, only for four stars, two hot EHes: HD\,124448 and PV\,Tel, and two C-poor hot EHes: V652\,Her and HD\,144941 where oberved Ne is significantly lower than the initial C + N + O + Ne limit,  F\,{\sc ii} detections are absent and hence, only the upper limits to the fluorine abundance are placed.

With the single exception of HD\,144941,  the C-poor EHe,  the N abundances of the EHe and RCB stars suggest an atmosphere dominated by gas seriously exposed to H-burning such that the initial C, N and O expected from the Fe abundance has been converted to N through the CNO-cycles.  The measured  Ne abundances of the majority of the  hot EHes and the inferred (that is observed non-LTE corrected) Ne abundances for the cool EHes and RCBs indicate that the Ne as $^{22}$Ne was most likely produced with near 100\% efficiency by $\alpha$-captures from the N by $\alpha$-captures in gas previously exposed heavily to H-burning.  These episodes of  H-burning and (partial) He-burning can not have occurred in the same gas:  Ne synthesis destroys the N and, in addition,  all Ne-rich stars have abundant O.  This juxtaposition of abundant N and Ne may be a pointer to distinct regions of nucleosynthesis and, perhaps, to a previous history as a binary system.

\begin{figure*}
\center
\includegraphics[scale=0.7]{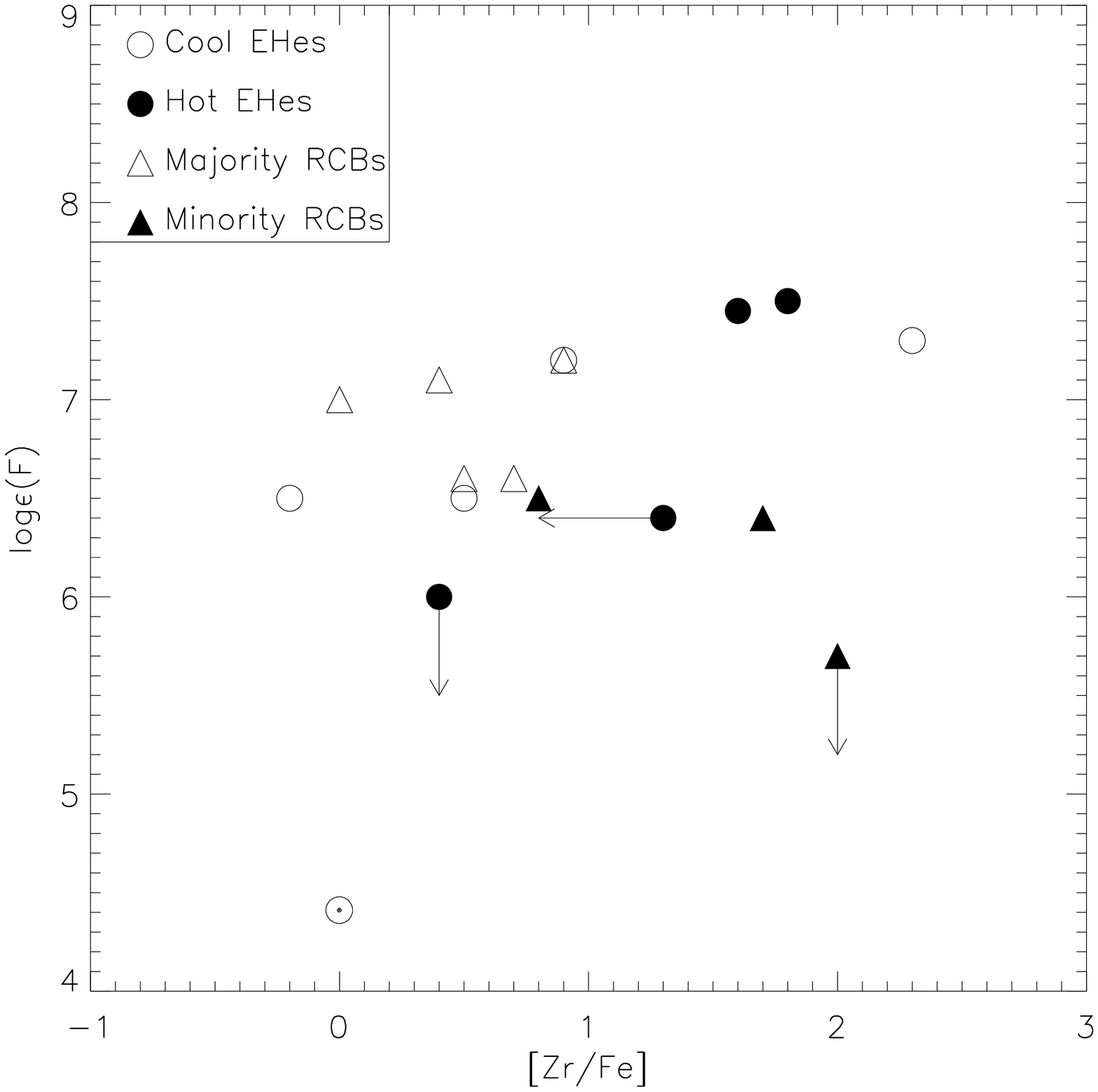}
\caption{\small{Observed  log$\epsilon$(F) with [Zr/Fe] for EHes and RCBs. The encircled dot symbol represents the sun. }}\label{fig.18}
\end{figure*}

Heavy elements offer another signature of nucleosynthesis, namely the $s$-process. \citet{Asplund2000} noted the overabundance of $s$-process elements in some RCB stars. Overabundances of Zr are found  for some hot EHes but in so few stars that a dependence of F on the $s$-process can not be determined  Table \ref{Table.5}. To the hot EHe sample, we add RCB Zr abundances from \citet{Asplund2000} and \citet{Hema2017}.  Zr for a sample of cool and hot EHes are from \citet{Pandey2006}'s analysis of {\it HST} ultraviolet spectra and Zr for other cool EHes are from \citet{Pandey2001} and \citet{Pandey2006r}. The full sample with both F and Zr abundances are shown in Figure \ref{fig.18}.  Severe $s$-process enrichment is certainly present among these H-deficient stars: [Zr/Fe] can exceed $+2$ but there are also stars lacking in detectable enrichment (i.e., [Zr/Fe] = 0). There is no obvious correlation between the F abundance and [Zr/Fe].

\section{Double white dwarf mergers and the fluorine abundance}

Until recently, two scenarios were in competition to explain the sequence EHe -- RCB -- HdC: the double-degenerate (DD) and the final-flash (FF) model. 

In the FF model, a late or final He-shell flash occurs in a post-AGB star, a star on the white dwarf cooling track, and converts the star to a H-poor cool luminous star (i.e., a HdC or RCB star) which then evolves back at about constant luminosity (i.e., as a EHe star) to the white dwarf cooling track \citep{Iben1983,Herwig2001}. Nucleosynthesis occuring during and following the He-shell flash shows that a H-poor supergiant may result with features of the composition characteristic of EHe, RCB and HdC stars but it has proven difficult to account for the key features namely, their low $^{16}$O/$^{18}$O ratios and their remarkable F overabundances \citep{Clayton2007,Pandey2006f,Pandey2008,Hema2017}. The FF model may yet be shown to account for other kinds of H-poor stars \citep{Pandey2011} such as V4334\,Sgr (Sakurai's object) \citep{Pandey2008}.

In the DD scenario, two white dwarfs merge. In the more favored version of the DD scenario, a He white dwarf is consumed by a more massive C-O white dwarf. In the less favoured version, two He white dwarfs merge. Population synthesis show that CO+He white dwarf binaries are much more likely than He+He white dwarf binaries. Neither version can account for the exceptional $^{18}$O abundances in HdC and RCBs and/or the extraordinary F abundances in RCBs and EHes without  episodes of nucleosynthesis accompanying the immediate phase of the merger and/or the post-merger phase. Clearly, the final compositions of the resulting  single H-deficient stars are likely to depend on the type of the merger — CO+He or He+He —  and on details of the stars (masses, compositions, etc.) comprising the close white dwarf binary which by loss of gravitational energy merges.  A merger results in a matter of minutes in a complex system comprising the core of the more massive white dwarf (i.e., the C-O white dwarf in the C-O+He system) surrounded by a  very hot corona (wonderfully dubbed `the shell of fire') inside a rapidly rotating disk. The disrupted less massive white dwarf is the principal contributor to the corona which may also receive mass from the more massive white dwarf. The less massive white dwarf is the principal contributor to the disk from which He-rich material is accreted by the central star on a slow time scale, say 10$^4$ -10$^5$ yr. After the central star has accreted sufficient material, He-shell burning commences  and the star's envelope expands to become a cool supergiant, that is a RCB or a HdC supergiant. The supergiant’s surface  composition is determined by the compositions of the merging white dwarfs, the nucleosynthesis occurring in the initial brief coronal  phase of the merger and in the supergiant’s, He-shell burning phase and by the (complex) physics of the whole merger process. Simulations of the merger and post-merger phases have yet to reach the finality to which the commendation `\textit{ab initio}' may be attached appropriately.

Our focus here is on published calculations of the DD scenario and their ability to match the observed F abundances of these H-deficient stars  and, in general, their  overall compositions including the remarkably low $^{16}$O/$^{18}$O ratios of HdC and RCB stars \citep{Clayton2007}, whose discovery has stimulated much of the theoretical work on these H-deficient stars.  For detailed descriptions of the various theoretical calculations one should consult the original papers. Our principal comparisons are with predictions for  CO+He white dwarf mergers provided by \citet{Lauer2019} and by \citet{Menon2013} and \citet{Menon2019}. \citet{Lauer2019} provide commentary on other calculations of CO+He white dwarf mergers including  \citet{Longland2011},  \citet{Zhang2014} and \citet{Menon2013} and \citet{Menon2019}.

\citet{Lauer2019} report on modeling of CO+He white dwarf mergers for stars initially of solar composition, i.e., [Fe] = 0. Most simulations consider a 0.55$M_\odot$ CO white dwarf leading to a post-merger mass of 0.8$M_\odot$. Predicted abundances for their principal product of a merger labelled A1 are summarized in their Figure 6. Model A1 deserves a fair pass against the observations of the $^{16}$O/$^{18}$O ratios and the F abundances of HdC, RCB and EHe stars extrapolated to [Fe] $\sim 0$.  Fluorine as synthesized in the hot corona is about a factor of three less than observed.  Observed C abundances are slightly under predicted. Model A1 also underpredicts the N abundance. Neon, as $^{22}$Ne , is predicted to be overabundant at the surface but quantitative estimates are not provided. For other elements — Na to Ti — observed and predicted abundances match quite well. Lithium production occurs in the A1 model providing a Li abundance at about the level seen in those few RCBs exhibiting Li.

Menon et al. in their 2013 paper considered four mergers for stars with initially [Fe]=0 and in their 2019 paper extended their study to merging stars with [Fe]=$-$1.4, and, thus, spanned the [Fe] range of the observed EHe and RCB stars.  Predicted surface compositions for [Fe]=0 and $-$1.4 with regards to $^{16}$O/$^{18}$O and F match observations quite well. The  predicted F enrichment reproduces the observed F abundances.  \citet{Menon2013} note that one source of F is in the He-burning shell of the post merger star where $^{13}$C$(\alpha,n)^{16}$O serves as a neutron source and $^{14}$N is both a neutron poison and a F source: $^{14}$N$(n,p)^{14}$C$(p,\gamma)^{15}$N$(\alpha,\gamma)^{19}$F. Predicted C and O abundances exceed observations, an issue discussed by \citet{Menon2019}. Note that the reported observed [O] abundances in Figure 5 of \citet{Menon2019} are overestimated by abount 0.7 dex. Predicted N abundances match observations quite well. In the hot corona, neutrons are released in some models and enrichment of $s$-process heavy elements predicted. Observed $s$-process enhancements are found in some EHes and RCBs, see Figure \ref{fig.18}. Predictions for Zr roughly match the observed maximum [Zr/Fe] for the models which release neutrons. The range of light $s$-process enhancements predicted  approximately matches the range of [Zr/Fe] shown in Figure \ref{fig.18}. However, one simulation experiencing  severe Zr enrichment also predicted substantial enrichment of heavy $s$-procss elements such as Ba and La, i.e., [Ba/Fe] = 4 which is not observed. Minor changes to the Na and Al abundances were predicted primarily as a result of proton captures.  Explicit predictions of the surface Ne abundances were not given but  appreciable synthesis of neon as $^{22}$Ne occurs in these models. In all published simulations the abundant isotope of neon is $^{22}$Ne not the commonly abundant $^{20}$Ne isotope.\footnote{Although the isotopic wavelength shifts of Ne\,{\sc i} and Ne\,{\sc ii} lines may be measurable with precision in the laboratory,  differentiation between the two Ne isotopes in a spectrum of a EHe will not be a trivial matter.  A catalog of isotopic shifts between $^{20}$Ne and $^{22}$Ne for Ne\,{\sc i} lines is given by \citet{Ohayon2019} and for Ne\,{\sc ii} lines by \citet{Oberg2007}. The maximum shift for our selection of lines for Ne\,{\sc i} is about 1.5 km s$^{-1}$ and for Ne\,{\sc ii} is about 2.5 km s$^{-1}$. With a careful selection of comparison lines around the Ne\,{\sc i} \&  Ne\,{\sc ii} lines it may be possible to show that the stellar Ne lines share the radial velocity of the star provided that the $^{22}$Ne wavelengths are adopted. However, the stellar lines for have a FWHM of about 10-30 km s$^{-1}$ and line widths and velocities may differ according their depth of formation in the stellar atmosphere.}

Considering the complexity of the physics  and the variety of initial conditions for the two white dwarfs in the CO+He  merger, it seems fair to conclude that the DD scenario with CO+He white dwarf mergers as presently simulated provides an adequate account of the two principal abundance anomalies of RCB and EHe stars, namely the $^{16}$O/$^{18}$O ratios and F abundances, and the C, N, O and Ne abundances without introducing other anomalies that are not matched by observations.  It remains to examine if the alternative possibility of He+He white dwarf mergers may also account for the compositions of some RCB and EHe stars.  Noting that \citet{Zhang2012l} indicate that  production of RCB and EHe stars via the He+He channel may be 14-70 times smaller than from the CO+He channel,  H-deficient stars created by the He+He channel may be the exception among the observed population of RCB and EHe stars.

Simulations of He+He white dwarf mergers as an explanation for H-deficient stars appear to be limited to those by \citet{Zhang2012,Zhang2012l} who explored restricted ranges for the many parameters entering into the simulations.  \citet{Zhang2012l} considered the merger of two 0.4$M_\odot$ He white dwarfs for four metallicities from Z = 0.02 to Z = 0.0001. Predicted surface abundances of the resulting RCB and EHe stars were ``in partial agreement" with the observed abundances.  In particular, the models showed `a strong overabundance of F [relative to the initial F abundance]'  but not enough to fully agree with the observational data. The disagreement was a factor of 100 at [Fe] = $-$2 decreasing to a factor of about 20 at [Fe] = $-$1.   In these mergers, the F is synthesized by  $^{14}$N$(\alpha,\gamma)^{18}$F$(p,\alpha)^{15}$O$(\alpha,\gamma)^{19}$Ne$(\beta^+)^{19}$F.   Enrichment of $^{18}$O may be underpredicted too. Minor disagreements between prediction and observation are found for C, N, O and Ne. The RCBs and EHes are predicted to be C-rich not C-poor at all Z. Lithium, which is observed in a few RCBs, is not  predicted to be present in the merged star.

In \citet{Zhang2012}, four models of equal-mass pairs of He white dwarfs were followed:  total masses considered were 0.5, 0.6, 0.7 and 0.8$M_\odot$ for initial compositions Z = 0.02 and 0.001. An aim of the calculations was to examine the effect of `slow’  mergers (the accreted star forms a disk around the accreting star from which gas is  accreted at $10^{-5}$ $M_\odot$ yr$^{-1}$), `fast’ mergers corresponding to an accretion rate of $10^4$ $M_\odot$ yr$^{-1}$ for the remaining white dwarf and `composite’ mergers  in which about 50 per cent of the donor star’s mass is accreted rapidly and the remainder forms a disk from which gas is accreted at the `slow’ rate.   Predicted compositions for the merged  0.5 - 0.8 $M\odot$ stars were given for just $^{12}$C, $^{14}$N, $^{18}$O and $^{22}$Ne and comparisons with observed compositions were made with helium-rich hot subdwarfs and not  RCB and EHe stars. While F was not reported for these simulations, their relevance  to our determinations of F abundances may be the conclusion that composite mergers from these equal mass white dwarf pairs  show an appreciable C underabundance for combined masses below about 0.6$M_\odot$  with little change of N across the mass range of 0.5-0.8$M_\odot$ and thus the N/C ratio is predicted to increase as the mass decreases below about 0.65$M_\odot$. This prediction, as \citet{Zhang2012} note, likely accounts for the two classes of He-sdO stars: the N-rich with N/C $>> 20$ and the C-rich with N/C $\leq 0.1$. The same prediction may provide the latitude to account for the C-poor  V652 Her and HD 144941 with their different N/C ratios  but their masses would have to be between 0.6 and 0.7$M_\odot$. One might also put the RCB XX Cam in this narrow range.

\section{Concluding remarks}

With observed determinations of the compositions of the H-deficient stars -- HdC, RCB and EHe -- and theoretical simulations of the merger of a CO white dwarf with a He white dwarf -- the DD scenario --, 
the many decade mystery surrounding compositions of these stars has been resolved. In particular, the large F overabundances for hot EHe stars derived in this paper  and compatible with F abundances obtained previously for cool EHe  and RCB stars are thanks to detailed simulations, e.g., \citet{Menon2013,Menon2019,Lauer2019}, of the DD scenario known to be quantitatively expected. Indeed, the simulations account well for the observed chemical compositions of the HdC, RCB and EHe sequence including the remarkably low $^{16}$O/$^{18}$O ratio \citep{Clayton2007} which with the F overabundances are {\it the} outstanding abundance anomalies of these H-deficient stars.

In the future, observers will be challenged to refine the determinations of chemical composition by not only obtaining more accurate analyses for  elements previously studied but by searching for the small abundance
changes in the elements Na to Zn and in the $s$-process elements predicted by the available quantitative studies of white dwarf mergers. A major lacuna in the abundance analyses concerns the non-LTE formation of the F\,{\sc i} and F\,{\sc ii} lines but this gap in quantitative knowledge does not affect the conclusion the F overabundance in these H-deficient stars is enormous and can be only slightly affected by inclusion of non-LTE effects. 

On the theoretical side, exploration of the DD scenario should continue. Predicted abundances of light elements  should be tested more thoroughly  than hitherto against observed abundance ratios.  For example, the puzzles represented in Figure \ref{fig.17} deserve close scrutiny: How can N and Ne both  have the  abundance implied by total conversion of initial  C, N and O?  
Perhaps, the range of chemical compositions of HdC, RCB and EHe stars may be used to set constraints on the boundary conditions for participants in a merger and the physical conditions during and following the merger with ultimate hope of achieving \textit{ab initio} predictions for the family of H-deficient stars.

\acknowledgements
 AB and GP thank the staff at the IAO, Hanle and at the remote control station at CREST, Bengaluru for their assistance with observations. We thank Falk Herwig for exchanges about Ne abundances in white dwarf mergers. We also thank the anonymous referee for the kind and encouraging comments.


\end{document}